\documentclass[12pt,aps,showpacs,preprintnumbers,amsmath,amssymb,amsfonts,superscriptaddress,openany]{revtex4-1}
\usepackage[utf8]{inputenc}
\usepackage{amsfonts}
\usepackage{amsbsy}
\usepackage{url}
\usepackage{enumerate}
\usepackage[justification=raggedright]{caption}
\usepackage{bbm}
\usepackage{mathdots}
\usepackage{graphicx}
\usepackage{graphicx,color}
\usepackage{amssymb,amsmath}
\usepackage{slashed}
\usepackage{hyperref}
\usepackage{braket}
\usepackage{simplewick}
\usepackage[title]{appendix}
\usepackage{caption}

\usepackage{ascmac}
\usepackage{latexsym}
\usepackage{pifont}          
\usepackage{bm}
\newcommand{\be}{\begin{eqnarray}}
\newcommand{\ee}{\end{eqnarray}}

\newcommand{\lcal}{\mathcal{L}}
\newcommand{\fcal}{\mathcal{F}}
\newcommand{\acal}{\mathcal{A}}
\newcommand{\p}{\partial}

\begin{document}

\preprint{YGHP-19-02}

\title{
Localization of gauge bosons and the Higgs mechanism on topological solitons in higher dimensions
}

\author{Minoru~Eto}
\affiliation{Department of Physics, Yamagata University, Kojirakawa-machi 1-4-12, Yamagata, Yamagata 990-8560, Japan}
\affiliation{Research and Education Center for Natural Sciences, Keio University, 4-1-1 Hiyoshi, Yokohama, Kanagawa 223-8521, Japan}

\author{Masaki~Kawaguchi}
\affiliation{Department of Physics, Yamagata University, Kojirakawa-machi 1-4-12, Yamagata, Yamagata 990-8560, Japan}

\begin{abstract}
We provide complete and self-contained formulas about
localization of massless/massive Abelian gauge fields on topological solitons in
generic $D$ dimensions via a field dependent gauge kinetic term.
The localization takes place when a stabilizer (a scalar field) is condensed 
in the topological soliton. We show that the localized gauge bosons are massless
when the stabilizer is neutral. On the other hand, they become massive for
the charged stabilizer as a consequence of interplay between
the localization mechanism and the Higgs mechanism.
For concreteness, we give two examples in six dimensions. The one is domain wall
intersections and the other is an axially symmetric soliton background.
\end{abstract}

\maketitle


\section{Introduction}
\label{sec:intro}

The idea that our world is a four-dimensional hyper surface (a 3-brane) in 
higher-dimensional spacetime has been widely studied for decades.
Such models are called brane-world models. 
By utilizing  {\it geometry} of the extra dimensions, 
the brane-world models can solve
many unsatisfactory problems of the Standard Model (SM), as provided by
the seminal works \cite{ArkaniHamed:1998rs,Antoniadis:1998ig,
Randall:1999ee,Randall:1999vf}.

A conventional setup of the brane-world models
is that extra dimensions are prepared as a compact 
manifold/orbifold. Namely, our four-dimensional spacetime is 
treated differently from the extra dimensions. 
Furthermore, 3-branes are introduced by hand
as non-dynamical objects which are infinitely thin.
In order to make the models more natural, we can harness {\it topology} 
of extra dimensions.
The idea is quite simple \cite{Rubakov:1983bb}: Dynamical
compactification of the extra dimensions and 
dynamical creation of 3-branes originate in a spontaneous 
symmetry breaking in a vacuum. Namely, it gives rise to a topologically stable soliton/defect 
of a finite width.
The topology ensures not only stability of the brane but also 
the presence of chiral matters localized on the brane 
\cite{Jackiw:1975fn,Rubakov:1983bb}.
In addition, graviton can be trapped \cite{Cvetic:1992bf,DeWolfe:1999cp,
Csaki:2000fc,Eto:2002ns,Eto:2003bn,Eto:2003ut}.

In contrast, localizing massless gauge bosons, especially non-Abelian 
gauge bosons, is quite difficult. There were many works so far 
\cite{Dvali:2000rx, Kehagias:2000au, Dubovsky:2001pe, 
Ghoroku:2001zu,Akhmedov:2001ny, Kogan:2001wp, Abe:2002rj, 
Laine:2002rh, Maru:2003mx, Batell:2006dp, Guerrero:2009ac, 
Cruz:2010zz, Chumbes:2011zt, Germani:2011cv, Delsate:2011aa, 
Cruz:2012kd, Herrera-Aguilar:2014oua, Zhao:2014gka, Vaquera-Araujo:2014tia,
Alencar:2014moa,Alencar:2015awa,Alencar:2015oka,Alencar:2015rtc,Alencar:2017dqb}. 
However, each of these has some advantages/disadvantages and 
there seems to be only little universal understanding. 
Then a new mechanism utilizing a field dependent gauge kinetic 
term (field dependent permeability)  
\be
- \beta(\phi_i)^2 F_{MN}F^{MN}, \quad (M,N=0,1,2,3,4,\cdots,D-1),
\label{eq:f2}
\ee
came out in Ref.~\cite{Ohta:2010fu} where $\phi_i$ are scalar fields which we
call {\it stabilizer} throughout this paper. 
This is a semi-classical realization of the confining phase  
\cite{ArkaniHamed:1998rs,Dvali:1996xe,Kogut:1974sn,Fukuda:1977wj, 
Luty:2002hj, Fukuda:2009zz, Fukuda:2008mz}.
Recently, one of the author and the collaborators 
have tried to improve the brane-world models with 
topological solitons by using (\ref{eq:f2}) \cite{Arai:2012cx,
Arai:2013mwa,Arai:2014hda,Arai:2016jij,Arai:2017lfv,Arai:2017ntb,
Arai:2018rwf,Arai:2018uoy,Arai:2018hao}.
Brief highlights of the results are the following: 
Ref.~\cite{Arai:2017ntb} provided the geometric 
Higgs mechanism which is the conventional Higgs mechanism driven 
by the positions of multiple domain walls in an extra dimension,
similarly to D-branes in superstring theories.
Then the geometric Higgs mechanism was applied to $SU(5)$ Grand Unified Theory  
in Ref.~\cite{Arai:2017lfv}. However, in these early works, treatment of gauge fixing
was partially unsatisfactory. 
In order to resolve this point, we have developed analysis 
by extending the $R_\xi$ gauge under spatially modulated backgrounds
in any spacetime dimensions $D$ \cite{Arai:2018rwf}.
Then, we gave a new attempt in Ref.\cite{Arai:2018uoy} that the SM Higgs field
plays a role of the stabilizer $\phi_i$ in a $D=5$ model.
The latest development along this direction is that the same localization mechanism
can be applied not only on vector fields but also scalars and tensors in Ref.~\cite{Arai:2018hao}.
Another group also recently studied the SM in a similar model 
with $\beta^2$ taken as a given background in $D=5$ 
\cite{Okada:2017omx,Okada:2018von,Okada:2019fgm}.

The purpose of this paper is to provide complete and self-contained formula about
localization of massless/massive Abelian gauge fields on topological solitons in
generic $D$ dimensions via the field dependent gauge kinetic term (\ref{eq:f2}).
The basic template is a domain wall in a $D=5$ model with stabilizers $\phi_i$ which
are neutral to the would-be localized gauge fields. Such models have been studied 
in Ref.~\cite{Ohta:2010fu,Arai:2012cx,
Arai:2013mwa,Arai:2014hda,Arai:2016jij,Arai:2017lfv,Arai:2017ntb}.
We firstly reanalyze it with the extended $R_\xi$ gauge to provide more complete arguments 
in Sec.~\ref{sec:2}.
Then, there are two directions for extending the basic template. The one is to go to higher dimensions
$D \ge 6$, and the other is to make the neutral stabilizers $\phi_i$ 
charged to the would-be localized gauge fields.
The former direction has been investigated in Ref.\cite{Arai:2018rwf} which we will give a review
with several improvements especially on divergence-free parts of extra-dimensional components
of the gauge fields in Sec.~\ref{subsec:neutral_D}. 
The latter extension that localization of gauge fields by
charged stabilizers was applied to the SM in $D=5$ dimensions \cite{Arai:2018uoy}.
In Ref.~\cite{Arai:2018uoy}, the charged stabilizer is nothing but the SM Higgs field.
Therefore, the Higgs field plays three roles: breaking of the electroweak symmetry,
generating fermion masses, and localizing the electroweak gauge bosons on the domain wall.
Since details on gauge field localization by the charged stabilizer were not given 
in Ref.~\cite{Arai:2018uoy}, in this work we will give a complete formula in details 
for accounting localization mechanism which occurs together with the Higgs mechanism
in Sec.~\ref{sec:charged_D5}.
Finally, we proceed to unify the two directions into the most generic models in $D\ge 6$
with charged stabilizers. 
In general, fluctuations are mixed each other under a soliton background,
so that it is not easy to distinguish which field is
physical or unphysical. To overcome this difficulty, we will carefully analyze the small fluctuations
in the extended $R_\xi$ gauge in Sec.~\ref{sec:Charged_Stab_D}, 
and succeed in cleaning up related Lagrangian up to quadratic
order of the fluctuations, and providing a simple and compact formula. With the
formula at hand, it becomes easy for us to obtain physical mass spectra 
appearing in a low energy effective theory on topological solitons  for generic models.
To be concrete, we give examples in Sec.~\ref{sec:example}. The first example is 
an intersection of two normal domain walls in $D=6$ dimensions. We will find an extremely
good approximation which allow us to compute analytically mass spectra localized on an
intersecting point. In the second example, we take an axially symmetric background on the assumption
that the topological soliton is a vortex-type soliton in $D=6$. 

For both cases that the stabilizer
is neutral and/or charged, our formula provides us a transparent road 
to obtain the physical mass spectra, and to specify the lightest
degrees of freedom in the low energy effective theory. We will figure out that the lightest fields
are massless four-dimensional 
gauge bosons and Nambu-Goldstone fields for the neutral stabilizer. There are infinite
tower of Kaluza-Klein (KK) modes but there is a large mass gap (of order inverse soliton width)
between the low-lying mode  and the higher KK modes.
Once we replace the neutral stabilizer by charged one, the Higgs mechanism occurs together with
localization of the gauge fields. As a result, the lowest mass mode becomes the massive four-dimensional
gauge field, and the heavy KK modes follow with the large mass gap. We should stress that
the scalar fields originated from the extra-dimensional components of the gauge fields are always above
the large mass gap, so that they do not
supply any low-lying physical degrees of freedom.

The organization on the paper is the following. 
The basic template of the localization mechanism with neutral stabilizers in $D=5$
is given in Sec.~\ref{sec:2}. The extension of Sec.~\ref{sec:2} with the neutral stabilizers
in higher dimensions $D\ge5$ is explained in Sec.~\ref{subsec:neutral_D}.
Sec.~\ref{sec:charged_D5} is devoted to explain
the other generalization of Sec.~\ref{sec:2} with charged stabilizers in $D=5$.
Then, the most generic models with charged stabilizers in higher dimensions $D\ge6$ is studied
in Sec.~\ref{sec:Charged_Stab_D}. Several concrete examples in $D=6$ are given in Sec.~\ref{sec:example},
and concluding remarks are given in Sec.~\ref{sec:conclusion}.

\section{Localization by a neutral stabilizer in $D=5$}
\label{sec:2}

\subsection{The model}
We consider the following Lagrangian in non-compact five dimensions,
\begin{align}
\lcal(H,T,{\cal A}_M) &= \lcal_{\rm g}(H,{\cal A}_M) + {\cal L}_{\rm s}(H,T),\label{eq:Lag_full}\\ 
\lcal_{\rm g}(H,{\cal A}_M)  &= - \beta ( H ) ^{2} \fcal _{MN} \fcal ^{MN}\label{lagrangian_0},
\end{align}
where $M,N = 0,1,2,3,4$, and $H$ is a complex scalar field, $T$ is a real scalar field,
and ${\cal A}_M$ is a $U(1)$ gauge field.
The field strength is given by
\be
\fcal_{MN} = \partial_{M} \acal_{N} - \partial_{N} \acal_{M}. 
\label{eq:cov_der}
\ee
For ${\cal L}$ to be real, $\beta$ should be a real functional of $H$. The scalar Lagrangian
${\cal L}_{\rm s}(H,T)$ depends on $H$ and $T$ as
\be
{\cal L}_{\rm s}(H,T) = \p_MH\p^MH^* + \p_MT\p^MT - V(H,T).
\label{eq:tilde_Lag}
\ee
As we will see below, $T$ is responsible for generating a non-trivial soliton background
configuration, and $H$ plays an important role for localizing gauge fields on the soliton.
We will refer to $H$ as a {\it stabilizer}. 
In this section, we will also assume that both $H$ and $T$ are 
neutral under the gauge transformation
associated with the gauge field ${\cal A}_M$; 
Namely, they do not interact with the gauge field ${\cal A}_M$ through conventional
covariant derivatives. Nevertheless, the neutral stabilizer $H$ can couple to ${\cal A}_M$
through the factor $\beta^2$ in front of the ${\cal F}^2$ term. 
Clearly, if $\beta$ is a constant, the Lagrangian reduces to
a conventional minimal one.
Note that 
$\beta^2$  can be interpreted as an extension of inverse square
of a gauge coupling constant which can be a function of the stabilizer $H$.

The Euler-Lagrange equations of the model read
\be
&&\p_M\p^M H + \frac{\p V}{\p H^*} =
-2\beta\frac{\p\beta}{\p H^*} {\cal F}_{MN}{\cal F}^{MN},\\
&&\p_M\p^M T + \frac{\p V}{\p T} = 0,\\
&&\p_M\left(\beta^2{\cal F}^{MN}\right) =0.
\ee
Clearly, ${\cal A}_M=0$ solves the last equation. Then, we 
are left with the first two equations with zeros at the right hand sides.
Hereafter, we will assume that $H$ and $T$ take a nontrivial background 
configuration depending only on the extra-dimensional
coordinate $y=x^4$. Such configurations generically appear 
as a topological soliton when
a discrete symmetry is spontaneously broken.

\subsection{An illustrative example}
\label{sec:5D_simplest}

Before describing a generic model, let us make a pause for illustrating
relevant phenomena through 
one of the simplest example in $D=5$. Let us consider the scalar potential
\be
V = \Omega^2 |H|^2 + \lambda^2 \left(T^2+|H|^2-v^2\right)^2.
\label{eq:V_example}
\ee
The model has the $U(1)$ global symmetry $H \to e^{i\alpha} H$, and
the $\mathbb{Z}_2$ symmetry $T \to -T$. When $\Omega^2 > 0$, there are two discrete vacua
$(H,T) = (0,\pm v)$. Thus, the $\mathbb{Z}_2$ symmetry is spontaneously broken, which
gives rise to a topologically stable domain wall:
It is straightforward to verify that the analytic domain wall solution is given by
\be
T_0=v \tanh \Omega (y-y_0),\quad
H_0= e^{i\alpha_0} v_H\, {\rm sech}\, \Omega (y-y_0),\qquad
\left(\Omega < \lambda v \right),
\label{eq:DW_sol1}
\ee
where $y_0$ and $\alpha_0$ are real moduli parameters, and we defined
$v_H \equiv \sqrt{v^2 - \frac{\Omega^2}{\lambda^2}}$.
The $\mathbb{Z}_2$ symmetry is recovered at the center of domain wall $y=y_0$, whereas
the unbroken symmetries at the vacua, namely the translational symmetry and the $U(1)$
global symmetry, are spontaneously broken only near the domain wall.
This locally broken symmetries are responsible for the presence of the moduli parameters
under the Nambu-Goldstone theorem.
Another domain wall solution is known for $\Omega \ge \lambda v$ as
\be
T_0 = v \tanh \lambda v (y-y_0),\quad H_0 = 0,\qquad \left(\Omega \ge \lambda v\right).
\label{eq:DW_sol2}
\ee
Here, the global $U(1)$ symmetry is unbroken everywhere. Thus, there is only the translational
moduli parameter $y_0$.

Either the translational symmetry or the global $U(1)$ symmetry is broken 
in the vicinity of the domain wall whereas they are not broken at the vacua.
Therefore, we expect the corresponding zero modes are localized on the domain wall.
This can easily be verified by perturbing the background solution as
\be
T = T_0(y) + \tau(x^\mu,y),\quad H = 
\left\{
\begin{array}{ccl}
h_1 + i h_2 & \ \text{for}\ & \Omega \ge \lambda v \\
e^{i\vartheta(x^\mu,y)}\left\{H_0(y) 
+ h(x^\mu,y)\right\} & \ \text{for}\  & \Omega < \lambda v
\end{array}
\right.
\label{eq:fluctuations}
\ee
where $\tau$, $h_{1,2}$, $h$ and $\vartheta$ are real.
In the following, we will set $y_0 = \alpha_0 = 0$ just for ease of the notation.
Plugging these into the Lagrangian ${\cal L}_{\rm s}(T,H)$ and taking the terms
quadratic in the fluctuations 
(we will shortly take fluctuations of the gauge fields into account below), we find
\be
{\cal L}_{\rm s}^{(2)} = 
\left\{
\begin{array}{ccl}
- \vec\xi^{\,\dagger}
\left(\p_\mu\p^\mu + {\cal M}^2 \right)
\vec\xi & \ \text{for}\ & \Omega \ge \lambda v\\
- \vec\xi^{\,T}
\left(\p_\mu\p^\mu + {\cal M}^2 \right)
\vec\xi
-H_0^2\, \vartheta \left(\p_\mu \p^\mu - H_0^{-2}\p_yH_0^2\p_y\right)\vartheta
& \ \text{for}\ & \Omega < \lambda v
\end{array}
\right..
\label{eq:tilde_L_2}
\ee
where we defined
\be
\vec\xi = 
\left(
\begin{array}{c}
\tau \\ h_1 + i h_2
\end{array}
\right)\ \text{for}\ \Omega \ge \lambda v,\qquad
\vec\xi = 
\left(
\begin{array}{c}
\tau \\ h
\end{array}
\right)
\ \text{for}\ \Omega < \lambda v,
\label{eq:xi}
\ee
and
\be
{\cal M}^2 = 
\left(
\begin{array}{cc}
-\p_y^2+2\lambda^2\left(3T_0^2+H_0^2-v^2\right) & 4\lambda^2T_0H_0 \\
4\lambda^2T_0H_0 & -\p_y^2 + 2\lambda^2\left(T_0^2+3H_0^2-v^2\right)+\Omega^2
\end{array}
\right).
\label{eq:M^2}
\ee

The KK mass spectra for $\vec \xi$ can be obtained by expanding $\vec\xi$ by
eigenstate of ${\cal M}^2$ as
\be
\vec\xi(x^\mu,y) = \sum_{n=0}^\infty \vec \Xi^{(n)}(y) \xi^{(n)}(x^\mu),\quad
{\cal M}^2\, \vec\Xi^{(n)} = m_{\xi,n}^2\, \vec\Xi^{(n)}.
\ee
Now, it is clear that there always exists the translational zero mode: 
\be
{\cal M}^2 \vec\Xi^{\,(0)} = 0,\quad
\vec\Xi^{\,(0)} = 
\p_y\left(
\begin{array}{c}
T_0(y)\\
H_0(y)
\end{array}
\right).
\label{eq:translation_NG}
\ee
As expected, this is localized around the domain wall, which can easily be verified by
looking at $|\vec\Xi^{(0)}|^2 = T_0'{}^2 + H_0'{}^2$.
Note that the translational zero mode exists regardless of values of the parameters $\Omega,v,\lambda$.
\begin{figure}[t]
\begin{center}
\includegraphics[width=10cm]{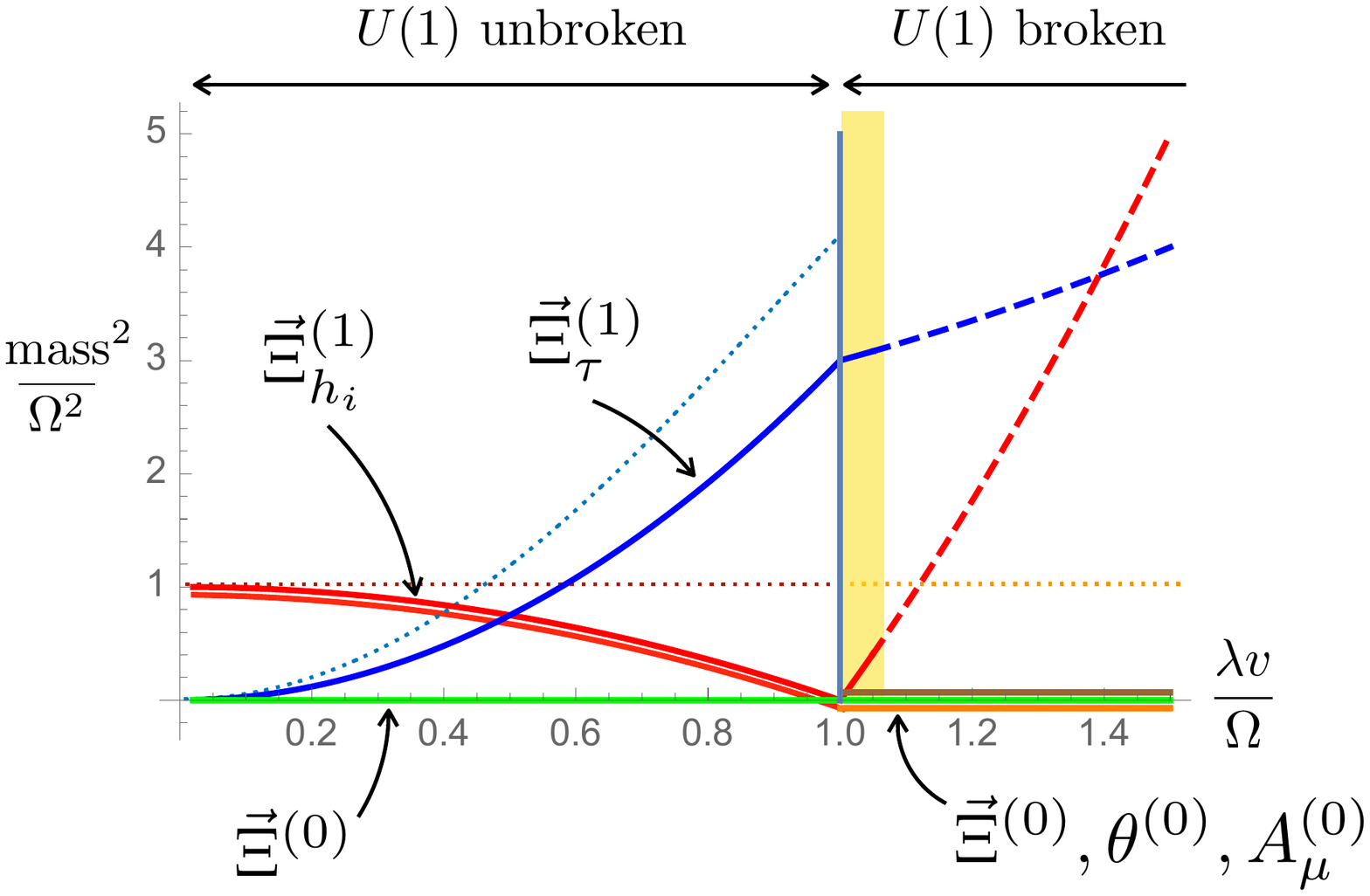}
\caption{Flows of mass eigenvalues of the localized states on the domain wall.
$\vec \Xi^{(0)}$ and $\theta^{(0)}$ correspond to the translational and $U(1)$ NG modes, respectively.}
\label{fig:mass}
\end{center}
\end{figure}
We can also derive several exact results for massive modes. 
Analysis for $\lambda v \le \Omega$ is easy since ${\cal M}^2$ is diagonal.
There are two orthogonal low lying mass eigenstates
\be
&&m_{\tau,1}^2 = 3\lambda^2 v^2,\quad 
\vec\Xi_{\tau}^{(1)} = \sqrt{\frac{3\lambda v}{2}}\left(
\begin{array}{c}
{\rm sech}\, \lambda v y~ \tanh\lambda v y\\
0
\end{array}
\right),\\
&&m_{h_i,1}^2 = \Omega^2-\lambda^2 v^2,\quad 
\vec\Xi_{h_i}^{(1)} = \sqrt{\frac{\lambda v}{2}}\left(
\begin{array}{c}
0\\
{\rm sech}\, \lambda v y
\end{array}
\right),\quad (i=1,2).
\ee
$m_{\tau,1}^2\Omega^{-2}$ is monotonically increasing function of
$\lambda v/\Omega$ whereas the degenerate masses $m_{h_i,1}^2\Omega^{-2}$ are monotonically decreasing
function. The two masses cross at $\lambda v/\Omega = 1/2$, see Fig.~\ref{fig:mass}.
The thresholds (the dotted lines in Fig.~\ref{fig:mass}) between localized discrete modes
and scattering continuum modes are $2\lambda v$ and $\Omega$ for $\tau$
and $h_i$, respectively.
At the critical point $\lambda v = \Omega$, $\vec\Xi_{h_i}^{(1)}$ becomes
exactly massless, whereas $\vec\Xi^{(1)}_\tau$ becomes heavy whose
mass is of order $\Omega$. There, the quadratic Lagrangian switches from the upper to the lower one
in Eq.~(\ref{eq:tilde_L_2}).
It is not easy to analytically compute mass eigenvalues for $\lambda v > \Omega$,\footnote{The
negative mass square $m^2_{h,1}$ implies that $H_0 = 0$ is unstable 
for $\lambda v > \Omega$. Indeed, $H_0 \neq 0$ for the solution in Eq.~(\ref{eq:DW_sol1}).}
since ${\cal M}^2$ is no longer diagonal. However, due to the continuity at $\lambda v = \Omega$, 
$\vec\Xi^{(1)}_\tau$ and $\vec\Xi^{(1)}_{h_i}$ should continuously be connected to corresponding degrees of
freedom for $\Omega > \lambda v$. 
Let us estimate the eigenvalues by treating $H_0$ as a small
perturbation.
To the first order of $\epsilon^2$ defined by
\be
\epsilon^2 \equiv \frac{\lambda^2v^2}{\Omega^2}-1, 
\ee
the mass eigenvalues are calculated as
\be
m_{\tau,1}^2\big|_{\Omega \lesssim \lambda v} &\simeq& 
\int^\infty_{-\infty} dy\, \vec\Xi^{(1)T}_\tau {\cal M}^2 \vec\Xi^{(1)}_\tau = 3\Omega^2 + \frac{4\Omega^2}{5}\epsilon^2,\\
m_{h,1}^2\big|_{\Omega \lesssim \lambda v} &\simeq& 
\int^\infty_{-\infty} dy\, \vec\Xi^{(1)T}_h {\cal M}^2 \vec\Xi^{(1)}_h =  
4\Omega^2 \epsilon^2.
\ee
Thus, the both eigenvalues grow up as $\lambda v/\Omega$ increasing, see the yellow band 
in Fig.~\ref{fig:mass}.

On the other hand, a linear combination of $\vec\Xi^{(1)}_{h_{1,2}}$ remains massless, and transforms to
the $U(1)$ NG mode. Indeed, 
$H_0 \propto {\rm sech}\,\Omega y$ never vanishes at finite $y$ for $\Omega < \lambda v$, so that
$\vartheta$ has the non minimal kinetic term.
To make the lower Lagrangian of Eq.~(\ref{eq:tilde_L_2}) be canonical,
let us redefine $\vartheta$ by
\be
\vartheta = \frac{\theta}{\sqrt2\,H_0}.
\label{eq:canonical_theta}
\ee
Then, the quadratic Lagrangian can be rewritten as
\be
{\cal L}^{(2)}_{\rm s} = - 
\vec\xi^{\,T}
\left(\p_\mu\p^\mu + {\cal M}^2 \right)
\vec\xi
- \frac{1}{2} \theta \left( \p_\mu\p^\mu +
Q_y^\dagger Q_y
\right)
\theta,
\label{eq:tildeL_2}
\ee
with
\be
Q_y \equiv - \p_y + \frac{\p_yH_0}{H_0},\quad
Q_y^\dagger \equiv \p_y + \frac{\p_yH_0}{H_0}.
\label{eq:D_5}
\ee
The mass spectrum corresponds to the eigenvalues of the Hermitian operator $Q_y^\dagger Q_y$ as
\be
Q_y^\dagger Q_y q^{(n)} = m_{Q,n}^2 q^{(n)}.
\label{eq:DD_5d}
\ee
Since $Q_y^\dagger Q_y$ is positive semidefinite, the eigenvalues are $m_{Q,n}^2 \ge 0$.
The normalizable zero mode is uniquely given by
\be
m_{Q,0} = 0,\quad 
q^{(0)} = H_0(y).
\label{eq:NG_U(1)}
\ee
There are no other localized modes.
All the excited states are continuum modes and given by
\be
m_Q(k) = \sqrt{\Omega^2 + k^2}\,,\quad 
q(y;k) = \frac{e^{iky}}{m_Q(k)}\left(k + i \Omega \tanh\Omega y\right).
\label{eq:continuum_sech}
\ee
Thus, the mass gap between the zero mode (the $U(1)$ NG mode $q^{(0)}$) 
and the continuum scattering modes is $\Omega$.

The final piece of the fluctuation analysis is on the gauge fields.
Since the background gauge fields are zeros, let ${\cal A}_M$ itself be fluctuations.
A relevant part to quadratic order in the fluctuation ${\cal A}_M$ is
\be
{\cal L}_{\rm g}^{(2)} = - \beta_0^2 \left(\p_M{\cal A}_N - \p_N{\cal A}_M\right)^2,\quad
\beta_0(y) \equiv \beta (H_0(y)).
\ee
To illustrate essence, let us take one of the simplest example
\be
\beta(H) = \frac{|H|}{2\mu}
\quad \to \quad
\beta_0(y) = \frac{v_H}{2\mu}\,{\rm sech}\,\Omega y,
\label{eq:beta_example}
\ee
where $\mu$ is a constant for $\lambda v > \Omega$.
Since nothing interests happen for $\lambda v \le \Omega$, 
we will investigate the parameter region $\lambda v > \Omega$ in what follows.
To figure out mass spectrum of the gauge field, we first need to
fix the $U(1)$ gauge symmetry. To this end, we add the following to the quadratic Lagrangian
\be
{\cal L}_\xi = -\frac{2\beta^2}{\xi}\left(\p_\mu {\cal A}^\mu 
- \xi\frac{1}{\beta^2}\p_y \left(\beta^2{\cal A}_y\right)\right)^2.
\label{eq:gauge_fix}
\ee
Here, $\xi$ is a gauge fixing parameter. If $\beta$ is a constant, it reduces to a conventional
covariant gauge ${\cal L}_\xi = - \frac{2\beta^2}{\xi}(\p_M{\cal A}^M)^2$. We call Eq.~(\ref{eq:gauge_fix})
the extended $R_\xi$ gauge \cite{Arai:2018rwf,Arai:2018uoy,Arai:2018hao,Okada:2017omx,Okada:2018von,Okada:2019fgm}.
In terms of the canonically normalized gauge field defined by
\be
A_M \equiv 2\beta_0 {\cal A}_M,
\ee
the quadratic Lagrangian reads
\be
\left({\cal L}_{\rm g} + {\cal L}_\xi\right)\bigg|_{\text{quad}} &=& 
\frac{1}{2}A_\mu\left[\eta^{\mu\nu}\square - \left(1 - \frac{1}{\xi}\right)\p^\mu\p^\nu 
+ \eta^{\mu\nu}
Q_y^\dagger Q_y
\right] A_\nu \nonumber\\
&&-\, \frac{1}{2}A_y 
\left(\square  + \xi
Q_y Q_y^\dagger
\right) A_y\,.
\label{eq:L2_5D}
\ee
Thanks to the gauge fixing term in Eq.~(\ref{eq:gauge_fix}), $A_\mu$ and $A_y$ are not mixed.
Hence, the KK mass spectra of $A_\mu$ and $A_y$ correspond to eigenvalues of $Q_y^\dagger Q_y$
and $\xi Q_yQ_y^\dagger$, respectively. It is coincident for the particular choice of $\beta$
in Eq.~(\ref{eq:beta_example}) that the Schr\"odinger operators
for $\theta$ and $A_\mu$ are identical (As we will see below, they are different in generic models.).
Therefore, the KK mass spectrum of $A_\mu$ consists of 
$m_{Q,0} = 0$ for the localized mode and $m_Q(k) = \sqrt{\Omega^2+k^2}$ for the continuum scattering 
states with the momentum $k$. 
Fortunately, no further computations are needed for $A_y$, since
$Q_y^\dagger Q_y$ and $Q_yQ_y^\dagger$ operators have the same mass eigenvalues
except for a zero eigenvalue.
Therefore, the mass square for $A_y$ is given by $\xi m_Q(k)^2 = \xi \left(\Omega^2 + k^2\right)$. 
It is peculiar that they depend on
the gauge fixing parameter $\xi$. Since any observable should not depend on gauge, we conclude that
$A_y$ do not include any unphysical degrees of freedom.
As we will show below, one can understand they are eaten by $A_\mu$ to give longitudinal modes of
the massive KK modes of $A_\mu^{({\rm KK})}$.


\subsection{Generic models}
\label{sec:generic_D5}

So far, we have argued over the specific model with
the scalar potential (\ref{eq:V_example}) and $\beta$ linear in $H$ given in Eq.~(\ref{eq:beta_example}). 
For a generic scalar potential, the matrix ${\cal M}^2$ given in Eq.~(\ref{eq:M^2}) is  modified as
\be
{\cal M}^2 = \left(
\begin{array}{cc}
-\p_y^2 + \frac{\partial^2 V(T_0,H_0)}{\partial T_0^2} & \frac{\partial^2 V(T_0,H_0)}{\partial T_0\partial H_0}\\
\frac{\partial^2 V(T_0,H_0)}{\partial T_0\partial H_0} & -\p_y^2+\frac{\partial^2 V(T_0,H_0)}{\partial H_0^2}
\end{array}
\right).
\label{eq:M^2_D5}
\ee
Accordingly, the detail mass spectrum of the scalar fields are 
different from those for the simplest model considered above. However,
the presence of the translational NG mode is intact, and indeed the mode function
$\vec \Xi^{(0)}$ given in Eq.~(\ref{eq:translation_NG}) formally remains correct. 
Furthermore, the quadratic Lagrangian of the scalar sector given 
in Eq.~(\ref{eq:tildeL_2}) is also formally valid for the generic model.
Therefore, we again have the following for $H_0 \neq 0$
\be
{\cal L}^{(2)}_{\rm s} = - 
\vec\xi^{\,T}
\left(\p_\mu\p^\mu + {\cal M}^2 \right)
\vec\xi
- \frac{1}{2} \theta \left( \p_\mu\p^\mu +
Q_y^\dagger Q_y
\right)
\theta.
\ee
Here, although the profile $H_0$ itself 
is different from Eq.~(\ref{eq:DW_sol1}) in general, 
but the definition of the operators $Q_y$ and $Q_y^\dagger$ are same as Eq.~(\ref{eq:D_5}).

On the other hand, the gauge sector given in Eq.~(\ref{eq:L2_5D})
gets modified when we generalize the linear function
$\beta(H) \propto H$ to a generic function $\beta(H)$.
After a little computations, we get
\be
\left({\cal L}_{\rm g} + {\cal L}_\xi\right)\bigg|_{\text{quad}} &=& 
\frac{1}{2}A_\mu\left[\eta^{\mu\nu}\square - \left(1 - \frac{1}{\xi}\right)\p^\mu\p^\nu 
+ \eta^{\mu\nu}
D_y^\dagger D_y
\right] A_\nu \nonumber\\
&&-\, \frac{1}{2}A_y 
\left(\square  + \xi
D_y D_y^\dagger
\right) A_y\,.
\label{eq:L2_5D_2}
\ee
Compared to Eq.~(\ref{eq:L2_5D}), $Q_y$ and $Q_y^\dagger$ are replaced by
$D_y$ and $D_y^\dagger$ which are defined by
\be
D_y = - \p_y + \frac{\p_y\beta_0}{\beta_0},\quad
D_y^\dagger =  \p_y + \frac{\p_y\beta_0}{\beta_0}\,.
\label{eq:def_D}
\ee
Note that $D_y = Q_y$ and $D_y^\dagger = Q_y^\dagger$ hold when $\beta$ is linear in $H$.

To complete the analysis, let us expand $A_\mu$ 
and $A_y$ by eigenfunctions of $D_y^\dagger D_y$ and $D_yD_y^\dagger$, respectively.
To this end, let us introduce the eigenfunctions
\be
D_y^\dagger D_y d^{(n)}(y) = m_{D,n}^2 d^{(n)}(y).
\label{eq:D_y^dagD_y}
\ee
Similarly to $Q^\dagger_y Q_y$, $D_y^\dagger D_y$ is positive semidefinite, so that $m_{D,n}^2 \ge 0$.
The normalizable zero mode, if it exists, is unique. It is given by
\be
m_{D,0} = 0,\qquad d^{(0)}(y) = \beta_0(y).
\ee
Hereafter, we assume that $\beta_0(y) = \beta(H_0(y))$ is non zero and square integrable
\be
0<\int^\infty_{-\infty} dy\, \beta_0(y)^2 < \infty.
\label{eq:square_integrable}
\ee
We expand $A_\mu$ in terms of $d^{(n)}$ as
\be
A_\mu(x,y) &=& \sum_{n=0}^\infty A_\mu^{(n)}(x) d^{(n)}(y).
\label{eq:A_mu_decompose_5D}
\ee
Thus, we conclude that $A_\mu$ always has the unique normalizable massless mode $A^{(0)}_\mu$.

Similarly, we expand $A_y$ by the eigenfunctions of $D_yD_y^\dagger$
\be
D_yD_y^\dagger\, \tilde d^{(n)}(y) = \tilde m_{D,n}^2\, \tilde d^{(n)}(y).
\ee
As $m_{D,n}^2$, $\tilde m_{D,n}^2 \ge 0$ holds. The massless eigenfunction is given by
\be
\tilde d^{(0)}(y) = \beta_0(y)^{-1}.
\label{eq:tilde_d0}
\ee 
However, this is non normalizable since we always impose the square integrability condition
(\ref{eq:square_integrable}). As is well-known, for the massive modes with $n>0$, we have
the following relations
\be
\tilde d^{(n)}(y) = \frac{D_y d^{(n)}(y)}{m_{D,n}},\qquad
m_{D,n} = \tilde m_{D,n}.
\ee
Thus, $A_y$ is expanded as
\be
A_y(x,y) = \sum_{n\neq0}^\infty a^{(n)}(x) \tilde d^{(n)}(y) =
\sum_{n\neq0}^\infty a^{(n)}(x) \frac{D_y d^{(n)}(y)}{m_{D,n}}.
\label{eq:Ay_decompose}
\ee

Plugging these into Eq.~(\ref{eq:L2_5D_2}) and integrating it over $y$, 
we find the quadratic Lagrangian for the low energy effective theory
\be
\int dy
\left({\cal L}_{\rm g} + {\cal L}_\xi\right)\bigg|_{\text{quad}} = 
\sum_{n=0}^\infty L_{{\rm g}}^{(n)},
\ee
with
\be
\!\!\!\!\!\!\!\!\!\!\!\!   L_{{\rm g}}^{(0)} &=&
\frac{1}{2}A^{(0)}_\mu\left[\eta^{\mu\nu}\square - \left(1 - \frac{1}{\xi}\right)\p^\mu\p^\nu 
\right] A^{(0)}_\nu,\\
\!\!\!\!\!\!\!\!\!\!\!\!   L_{{\rm g}}^{(n>0)} &=& 
\frac{1}{2}A^{(n)}_\mu\left[\eta^{\mu\nu}\square - \left(1 - \frac{1}{\xi}\right)\p^\mu\p^\nu 
+ \eta^{\mu\nu} m_{D,n}^2
\right] A^{(n)}_\nu
- \frac{1}{2} a^{(n)} 
\left(\square  + \xi
m_{D,n}^2\right) a^{(n)}\,.
\label{eq:L_KK_expand_5d}
\ee
Clearly, $L_{\rm g}^{(0)}$ is a usual Lagrangian in the $R_\xi$ gauge (covariant gauge) 
for the massless Abelian gauge field
$A_\mu^{(0)}$ in four dimensions. Similarly, $L_{\rm g}^{(n>0)}$ is identical to a conventional 
Lagrangian in the $R_\xi$ gauge for
the Abelian gauge field $A_\mu^{(n>0)}$ which gets the mass $m_{D,n}$ as a result of the Higgs mechanism
by absorbing the scalar field $a^{(n>0)}$. Thus, $A_y$ does not provide any
physical degrees of freedom to the low energy effective action but is converted into 
longitudinal degrees of freedom
to $A_\mu^{(n>0)}$.

Before closing this section, for later convenience, let us describe the above results from
a slightly different view point.
We decompose $A_y$ into a divergence part and a divergence-free part as
\be
A_y = A_y^{\rm d} + A_y^{\rm df},
\ee
where the separation is done by the following projection operator
\be
P = D_y \left(D_y^\dagger D_y\right)^{-1} D_y^\dagger,\quad
A_y^{\rm d} = P A_y,\quad A_y^{\rm df} = (1-P)A_y.
\ee
Then, it holds
\be
D_y^\dagger A_y^{\rm df} = 0.
\label{eq:DydagAy}
\ee
Furthermore, we can easily show $D_y^\dagger A_y$ is orthogonal 
to the zero mode $d^{(0)}$ of $D_y^\dagger D_y$ as
\be
\int dy\, d^{(0)}D_y^\dagger A_y = \int dy\, (D_y d^{(0)})A_y = 0.
\label{eq:div_5D}
\ee
This ensures that the projection operator $P$ is well-defined. 
Now, we can rewrite the quadratic Lagrangian
of the gauge sector given in Eq.~(\ref{eq:L2_5D_2}) as
\be
\left({\cal L}_{\rm g} + {\cal L}_\xi\right)\bigg|_{\text{quad}} 
&=& 
\frac{1}{2}A_\mu\left[\eta^{\mu\nu}\square - \left(1 - \frac{1}{\xi}\right)\p^\mu\p^\nu 
+ \eta^{\mu\nu}
D_y^\dagger D_y
\right] A_\nu \nonumber\\
&&-\, \frac{1}{2}A_y^{\rm df} 
\square A_y^{\rm df}
- \frac{1}{2}A_y^{\rm d} 
\left(\square  + \xi
D_y D_y^\dagger
\right) A_y^{\rm d}\,.
\label{eq:5d_Lfluc_A}
\ee
Note that
the divergence-free condition (\ref{eq:DydagAy}) implies
\be
A_y^{\rm df} = \eta \beta_0^{-1},
\label{eq:divfree_5d}
\ee
with $\eta$ being a constant in $y$. Indeed,
this corresponds to the non-normalizable zero mode $\tilde d^{(0)}$
of $D_yD_y^\dagger$
given in Eq.~(\ref{eq:tilde_d0}).
Therefore, eliminating $\tilde d^{(0)}$ from $A_y$ in Eq.~(\ref{eq:Ay_decompose}) is nothing but
getting rid of the divergence-free part $A_y^{\rm df}$ form $A_y$.
Thus, eliminating the unphysical $A_y^{\rm df}$, the quadratic Lagrangian reduces to
\be
\left({\cal L}_{\rm g} + {\cal L}_\xi\right)\bigg|_{\text{quad}} 
&=& 
\frac{1}{2}A_\mu\left[\eta^{\mu\nu}\square - \left(1 - \frac{1}{\xi}\right)\p^\mu\p^\nu 
+ \eta^{\mu\nu}
D_y^\dagger D_y
\right] A_\nu \nonumber\\
&&
-\, \frac{1}{2}a
\left(\square  + \xi
D_y^\dagger D_y
\right) a\,,
\ee
where we defined
\be
a = \frac{1}{\sqrt{D_y^\dagger D_y}} D_y^\dagger A_y
= \frac{1}{\sqrt{D_y^\dagger D_y}} D_y^\dagger A_y^{\rm d}.
\label{eq:tilde_A_5D}
\ee
Now, we run into a direct correspondence between the mass square operators $D_y^\dagger D_y$ for $A_\mu$
and $\xi D_y^\dagger D_y$ for $a$.
This implies $a$ is the one which is eaten by $A_\mu^{(n>0)}$.
Thus, our statement becomes more solid than before: 
{\it The divergence-free part of $A_y$ is unphysical because
it diverges at the spatial infinity. 
The divergence part of $A_y$ is also unphysical
because it is absorbed by $A_\mu^{(n>0)}$.}

Since $D_y^\dagger A_y$ is orthogonal to $d^{(0)}$, $a$ is also orthogonal to $d^{(0)}$. 
Thus, we can expand $a$ by
the eigenfunction $d^{(n)}$ of $D_y^\dagger D_y$ as
\be
a(x,y) = \sum_{n>0}^\infty a^{(n)}(x) d^{(n)}(y).
\ee
This is perfectly consistent with Eqs.~(\ref{eq:Ay_decompose}) and (\ref{eq:tilde_A_5D}).
Comparing this with the decomposition of $A_\mu$ in Eq.~(\ref{eq:A_mu_decompose_5D}), it is quite
natural that the absorbing state $A_\mu^{(n>0)}$ and the absorbed state $a^{(n>0)}$ have the
same wave function $d^{(n>0)}$.

\begin{figure}
\begin{center}
\includegraphics[width=15cm]{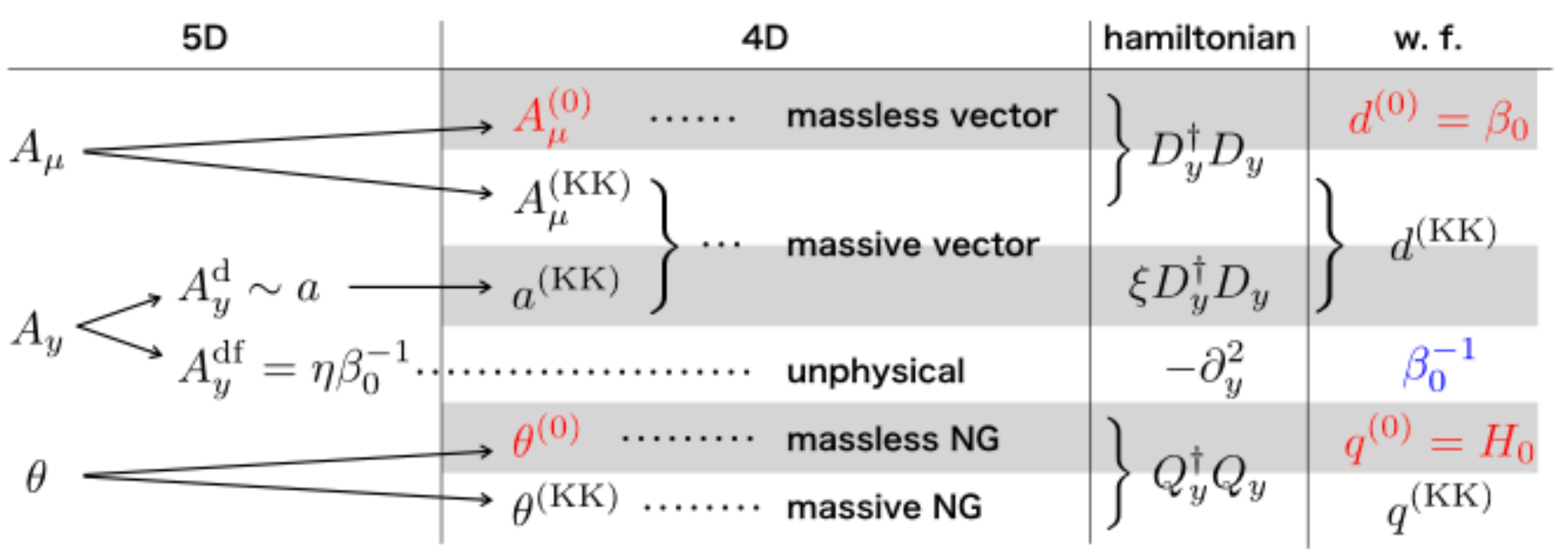}
\caption{A brief summary of the effective fields localized on 
the domain wall by a neutral stabilizer $H$ in five-dimensional models.}
\label{fig:summary5D}
\end{center}
\end{figure}

We briefly summarize the gauge sector with Fig.~\ref{fig:summary5D}.
There are only two massless states in the low energy effective theory in four dimensions:
the one is the gauge field $A_\mu^{(0)}$ and the other is the $U(1)$ NG field $\theta^{(0)}$.
There are infinite KK towers of $A_\mu^{({\rm KK})}$ (eating $a^{({\rm KK})}$), and $\theta^{({\rm KK})}$.
Several lower KK discrete modes would be localized on the domain wall, followed by the infinite KK bulk modes.
These spectra are dependent of the details of the model.  However, it is always true that the massless
states are gapped from the KK towers by the order of inverse of the domain wall width.

\section{Neutral stabilizers in higher dimensions}
\label{subsec:neutral_D}

Let us next extend the previous $5D$ models to generic $D$ dimensional models.
We study the Lagrangians which are formally same as those given in Eqs.~(\ref{eq:Lag_full})
and (\ref{lagrangian_0}), by reinterpreting the spacetime indices $M,N = 0,1,\cdots,D-1$.
In the following arguments, we will not specify a concrete model for the 
scalar part ${\cal L}_{\rm s}(H,T)$, as we did in Sec.~\ref{sec:generic_D5}.
Instead, we will assume that the Lagrangian ${\cal L}_{\rm s}(H,T)$ allows for the scalar field $T$ 
to give rise to a topological soliton
whose world-volume dimensions are four (codimension is $D-4$). For example, 
a domain wall in $D=5$, a vortex in $D=6$, a monopole in $D=7$, and so on.\footnote{
Here, $T$ symbolically stands for multiple scalar fields which are needed to form
a topological soliton.} Furthermore, we will assume that the neutral stabilizer $H$ locally condenses 
about the topological soliton as 
\be
H \to H_0(y) = \left\{
\begin{array}{ccl}
\text{non-zero} & \ \cdots\ & \text{around the topological soliton}\\
\text{zero} & \ \cdots\ & \text{far from the topological soliton}
\end{array}
\right.\,.
\ee
Hereafter, we will use $y$ to express the
extra dimensional coordinates $y = \{x^4,x^5,\cdots,x^{D-1}\}$.
Thus, the coefficient $\beta(H)$ in the gauge kinetic
term (\ref{lagrangian_0}) becomes a nontrivial 
function of the extra dimensional coordinates as
\be
\beta_0(y) = \beta(H_0(y)).
\ee
As a natural extension of Eq.~(\ref{eq:square_integrable}), 
it will turn out that the sufficient condition to $\beta_0$ 
for a massless $4D$ gauge field to be localized on the topological soliton is
the square integrability of $\beta_0$
\begin{align}
0 < \int d ^{D-4} y\, \beta_0 ^{2} < \infty. \label{square-integrability}
\end{align}
This implies that $ \beta $ is non-zero and quickly approaches to zero
at the extra-dimensional spatial infinity. 

\subsection{Useful formulae}
For later convenience, let us first correct some useful equations.
Let us introduce a differential operator $ D_{a} $ and
its adjoint operator $D_a^\dag$ by
\be
D_{a} = - \beta_0 \partial_{a} \dfrac{1}{\beta_0},\quad 
D_{a} ^{\dagger} = \dfrac{1}{\beta_0} \partial_{a} \beta_0,
\quad (a=4,5,\cdots,D-1).
\label{eq:D_a}
\ee
Clearly, these are natural extensions of $D_y$ and $D_y^\dagger$ 
given in Eq.~(\ref{eq:def_D}).
They satisfies the following algebra
\be
\left[ D_{a} , D^{\dagger}_{b} \right] = - 2 \left( \partial_{a} \partial_{b} \log \beta_0 \right) , \quad \left[ D_{a} , D_{b} \right] = \left[ D^{\dagger}_{a} , D^{\dagger}_{b} \right] = 0.
\ee
We will frequently encounter a self-adjoint operator defined by
\begin{align}
D_{a}^{\dagger} D_{a} = - \partial_{a}^{2} + \dfrac{( \partial_{a} ^{2} \beta_0 )}{ \beta_0 } \label{D2_operator},
\end{align}
where the sum on $a$ is implicitly taken. We will also use the following vector notation
\be
\vec D = \left(\begin{array}{c}
D_4\\
\vdots\\
D_{D-1}
\end{array}
\right),\qquad
\vec D^\dagger = \left(D_4^\dagger,\ \cdots,\ D_{D-1}^\dagger\right).
\ee
Clearly, $ \vec D^\dagger \vec D = D_{a}^{\dagger} D_{a} $ is positive semidefinite. 
Let $d^{(n)}$ be eigenstates of $\vec D^\dagger \vec D$ as
\be
\vec D^\dagger \vec D d^{(n)} = m_{D,n}^2 d^{(n)}.
\ee
Suppose $d^{(0)}$ be a normalizable 
eigenfunction. Then, we have
\be
0 
= \int d^{D-4}y\ d^{(0)} \vec D^\dagger \vec D\, d^{(0)}
= \int d^{D-4}y\ \left(D_a d^{(0)}\right)^2.
\ee
Therefore, $d^{(0)}$ must be annihilated by all $D_a$'s as
\be
D_a d^{(0)} = 0,\quad (a=4,5,\cdots).
\label{eq:d0_geneD}
\ee
There is a unique solution up to a normalization constant to these equations
\be
d^{(0)} = \beta_0.
\label{eq:D_0mode}
\ee
We now understand the square integrability condition (\ref{square-integrability}) is nothing 
but ensuring the normalizablity for the zero mode of $\vec D^\dagger \vec D$.
Namely, we here proved the existence and uniqueness of the normalizable zero
eigenfunction $d^{(0)}$ of $\vec D^\dagger \vec D$.
Note that the previous work \cite{Arai:2018rwf} also studied the zero mode $d^{(0)}$, 
where the uniqueness was only proved
for a radially symmetric background solution $\beta_0$ which is a function of $r = \sqrt{(x^a)^2}$.

Another self-adjoint operator will also take part in the following
arguments,
\be
D_aD_a^\dagger = - \p_a^2 + \frac{\left(\p_a^2\beta_0^{-1}\right)}{\beta_0^{-1}}.
\ee
Similarly to $\vec D^\dagger \vec D$, 
this operator is positive semidefinite. An obvious zero eigenstate $\tilde d^{(0)}$
is given by
\be
\tilde d^{(0)} = \frac{1}{\beta_0},
\ee
up to a constant. However, this is non-normalizable 
under the condition (\ref{square-integrability}). 
Thus, we conclude that there are no normalizable eigenstates
for zero eigenvalue of $D_a D_a^\dagger$.  

There is a useful corollary. Let $\{B_a(y)\}$ be a
set of $D-4$ non-singular and finite functions.
Then we find that $D_a B_a$ is orthonormal to $d^{(0)}$. This can be shown as follows:
\be
\int d^{D-4}y\, d^{(0)}D_a^\dagger B_a = \int d^{D-4}y\, \left(D_ad^{(0)}\right) B_a = 0,
\label{eq:orthogonality}
\ee
where we have used Eq.~(\ref{eq:d0_geneD}).

Finally, let us mention about $\vec D \vec D^\dagger$ which is ``super-partner'' of $\vec D^\dagger\vec D
= D_a^\dagger D_a$. Note that $\vec D \vec D^\dagger$ is a $(D-4) \times (D-4)$ matrix operator and
it is different form the one by one operator $D_a D_a^\dagger$. 
Zero eigenstates of $\vec D\vec D^\dagger$ are given by
\be
\vec {\tilde d}^{(0)} = \vec a \beta_0^{-1},
\ee
with $\vec a$ being a constant $D-1$ vector. 
This is non-normalizable under the condition (\ref{square-integrability}).
It is easy to show that non-zero eigenvalues
of $\vec D^\dagger \vec D$ and $\vec D \vec D^\dagger$ coincide as
\be
\vec D \vec D^\dagger (\vec D d^{(n)}) = \vec D (\vec D^\dagger \vec D d^{(n)})
= m_{D,n}^2 \vec D d^{(n)}.
\ee

\subsection{The quadratic Lagrangian}
With these preparations at hand, we are now ready to study mass spectra of the small fluctuations
about the soliton background.
We use the same notations for the scalar fluctuations 
$\tau(x^\mu,y)$, $\vartheta(x^\mu,y)$ and
$h(x^\mu,y)$ as Eq.~(\ref{eq:fluctuations}). Of course, here, we understand $y$ to represent the
extradimensional coordinates $y = \{x^a\}$. As in the five dimensional case, ${\cal A}_M=0$ for the
soliton background, and we again use ${\cal A}_M(x,y)$ itself for the small fluctuations.

First thing we have to do is to fix the $U(1)$ gauge symmetry.
To this end, similarly to the $5D$ case, we add the following
gauge fixing term
\begin{align}
\lcal_\xi = - \dfrac{2 \beta ^{2}}{\xi} 
\left(
\partial^{\mu} \acal_{\mu} + \xi \dfrac{1}{ \beta^{2} } \partial^{a} \left(\beta^{2} \acal_{a}\right)
\right)^2.
\label{eq:L_fix_D}
\end{align}
As before, 
we use canonically normalized fields $ A_{M} \equiv 2 \beta \acal_{M}$. 
Then, we find that the quadratic Lagrangian consists of two independent parts: The scalar part
and the gauge part as
\be
{\cal L}^{(2)} = {\cal L}_{\rm s}^{(2)} + (\lcal_{\rm g} + \lcal_\xi)\big|_{\text{quad}}.
\ee
The gauge part is given by
\be
(\lcal_{\rm g} + \lcal_\xi)\big|_{\text{quad}}
= \dfrac{1}{2} A_{\mu} \Gamma^{\mu\nu} A_{\nu}
+ \dfrac{1}{2} A_{a} \Gamma_{ab} A _{b},\label{lagrangian_0_4ex}
\ee
with
\be
\Gamma^{\mu\nu} &=& \eta ^{ \mu \nu } \Box - \left ( 1 - \dfrac{1}{\xi} \right ) \partial^{\mu} \partial^{\nu} + \eta ^{\mu \nu} \vec D^\dagger \vec D,
\label{eq:Gamma4}\\
\Gamma_{ab} &=& - \left(
\Box \delta _{ab} + {\cal H}_{ab} + \xi D_{a} D^{\dagger}_{b}
\right).\label{eq:Gamma5}
\ee
We defined
\be
{\cal H}_{ab} &\equiv& \delta_{ab} \vec D^\dagger \vec D 
- D_b^\dagger D_a,
\label{eq:Gamma_extra}
\ee
which satisfies the following identity
\be
{\cal H}_{ab}D_b = 0.
\label{eq:HD}
\ee
The scalar part for $H_0 \neq 0$ is given by
\be
{\cal L}^{(2)}_{\rm s} = - 
\vec\xi^{\,T}
\left(\square + {\cal M}^2 \right)
\vec\xi
-h_0^2\, \vartheta \left(\square - H_0^{-2}\p_a H_0^2\p_a\right)\vartheta\,,
\ee
with
\be
\vec\xi = \left(\begin{array}{c}
\tau\\
h
\end{array}
\right),\qquad
{\cal M}^2 = \left(
\begin{array}{cc}
-\p_a^2 + \frac{\partial^2 V(T_0,H_0)}{\partial T_0^2} & \frac{\partial^2 V(T_0,H_0)}{\partial T_0\partial H_0}\\
\frac{\partial^2 V(T_0,H_0)}{\partial T_0\partial H_0} & -\p_a^2+\frac{\partial^2 V(T_0,H_0)}{\partial H_0^2}
\end{array}
\right).
\label{eq:xi_M}
\ee

\subsubsection{The scalar part}
We can further rewrite the above scalar quadratic Lagrangian with respect to the canonically
normalized field $\theta$ defined in Eq.~(\ref{eq:canonical_theta}) as
\be
{\cal L}^{(2)}_{\rm s} = - 
\vec\xi^{\,T}
\left(\square + {\cal M}^2 \right)
\vec\xi
- \frac{1}{2} \theta \left( \square +
Q^\dagger_a Q_a
\right)
\theta,
\label{eq:Lag_2s_D}
\ee
where we defined the differential operators which are natural extension of $Q_y$ and $Q_y^\dagger$ by
\be
Q_a \equiv - \p_a + \frac{\p_aH_0}{H_0},\quad
Q_a^\dagger \equiv \p_a + \frac{\p_aH_0}{H_0}.
\label{eq:QD>6}
\ee
As in the five dimensional model, $Q_a$ and $D_a$ coincide if $\beta_0 \propto H_0$.
It is obvious that $Q^\dagger_a Q_a$ has a unique zero eigenstate
\be
q^{(0)} = H_0.
\ee
Thus we conclude that $\theta$ has the physical zero mode ($U(1)$ NG mode) and expanded as
\be
\theta (x,y) = H_0 \theta^{(0)}(x) + \sum_{n>0}^\infty q^{(n)}(y)\theta^{(n)}(x).
\ee

\subsubsection{The four dimensional component $A_\mu$}

The quadratic Lagrangian for $A_\mu$
\be
\dfrac{1}{2} A_{\mu} \Gamma^{\mu\nu} A_{\nu}
=\dfrac{1}{2} A_{\mu} 
\left(
\eta ^{ \mu \nu } \Box - \left ( 1 - \dfrac{1}{\xi} \right ) \partial^{\mu} \partial^{\nu} + \eta ^{\mu \nu} \vec D^\dagger \vec D
\right)
A_{\nu}
\label{eq:Lag_2Amu_D}
\ee
is a natural extension of Eq.~(\ref{eq:L2_5D_2}) by exchanging $D_y^\dagger D_y$
by $\vec D^\dagger \vec D$. We naturally expand the four dimensional components $A_\mu$
by the eigenfunctions of $\vec D^\dagger \vec D$ as
\begin{align}
A_{\mu}(x,y) = \sum_{n=0}^\infty 
A_{\mu} ^{(n)}(x) d^{(n)} (y).
\end{align}
As was proved in Eq.~(\ref{eq:D_0mode}),
the ground state is unique $d^{(0)} \propto \beta_0$ with
$m_0 = 0$, and there is a finite mass gap (about inverse of the background soliton width)
between the ground state and excited states.
Plugging this into the Lagrangian and integrating it in the extra dimensional coordinate $y$, 
we find the four dimensional effective Lagrangian
for $ A_{\mu}^{(n)}$ as
\begin{align}
\int d^{D-4} y\, \frac{1}{2}A_\mu \Gamma^{\mu\nu} A_\nu 
= \sum_{n = 0}^{\infty} \dfrac{1}{2} A_{\mu}^{(n)} \left [  \eta ^{ \mu \nu } \Box - \left ( 1 - \dfrac{1}{\xi} \right ) \partial^{\mu} \partial^{\nu} + \eta ^{\mu \nu} m_{n}^{2} \right ] A_{\nu}^{(n)}.
\label{eq:expand_A4_Dgene}
\end{align}

\subsubsection{The extra dimensional component $A_a$}
\label{sec:A_a_neutral_D}

Let us next turn to the extra dimensional components $A_a$ which provide additional scalar
fields to the low energy effective theory in four dimensions.
The multiple scalars $A_a$ make the quadratic Lagrangian more complicated
compared to the one in five dimensions. 
To make the matter clear, let us first decompose 
$A_a$ into a divergence part $A_a^{\rm d}$ and divergence-free parts $A_a^{\rm df}$  as
\begin{align}
A_{a}^{\rm d} = P_{ab} A_{b},\quad  
A_{a}^{\rm df} = ( \delta_{ab} - P_{ab}) A_{b}, \label{d,df_1}
\end{align}
with a projection operator 
\begin{align}
P_{ab} = D_{a} \left(\vec D^\dagger \vec D\right)^{-1} D^{\dagger}_{b}.
\end{align}
It satisfies the following equations
\begin{align}
P_{ab} P_{bc} = P_{ac} ,\quad
D^{\dagger}_{a} P_{ab} = D^{\dagger}_{b},\quad
P_{ab} D_{b} = D_{a}. \label{P_property}
\end{align}
With these at hand, it is straightforward to verify the divergence free condition should be satisfied
\be
D_a^\dagger A_a^{\rm df} = 0.
\label{eq:id0}
\ee
One immediately finds that $A_a^{\rm df}$ includes a component proportional to $\beta_0^{-1}$ as
\be
\vec A^{\rm df} = \beta_0^{-1} \vec\eta + \vec A^{\widehat{\rm df}},
\ee
with $\vec \eta$ being an arbitrary but constant $N$ vector, and
$\vec A^{\widehat{\rm df}}$ standing for rest component orthogonal to $\beta_0^{-1} \vec\eta$.
However, we should remove $\beta_0^{-1} \vec\eta$ since it diverges at the spatial infinity.
This is an extension of Eq.~(\ref{eq:divfree_5d}) in the $D=5$ case. In contrast to the $D=5$ case,
there still exist physical degrees of freedom
in $\vec A^{\widehat{\rm df}}$ in the higher dimensions.
From the definition of $A^{\rm d}_a$ given in Eq.~(\ref{d,df_1}) and the identity (\ref{eq:HD}),
it also follows
\be
{\cal H}_{ab} A^{\rm d}_b = 0.
\label{eq:id1}
\ee

Since the projection operator includes the inverse of $\vec D^\dagger \vec D$, 
one might worry whether it is 
well defined or not. However, it is always well defined  
because $D_a^\dagger A_a$ is orthonormal to the zero mode $d^{(0)}$ of $\vec D^\dagger \vec D$, see
the corollary in Eq.~(\ref{eq:orthogonality}).

Now, by using Eqs.~(\ref{eq:id0}) and (\ref{eq:id1}), we can rewrite the quadratic Lagrangian as
\be
\dfrac{1}{2} A_{a} \Gamma_{ab} A _{b}
= -\dfrac{1}{2} A^{\widehat{\rm df}}_a \left(\delta_{ab}\square + {\cal H}_{ab} \right) A^{\widehat{\rm df}}_b -
\dfrac{1}{2} A_a^{\rm d} \left(\delta_{ab}\square +  \xi D_aD_b^\dagger\right) A_b^{\rm d},
\label{eq:extraA_Lag_geneD}
\ee
The divergence and the divergence-free parts are decoupled.


\paragraph{The divergence part}

Let us first examine the divergence part.
The divergence part $A_a^{\rm d}$ essentially include only one independent
degree of freedom.
Similarly to the five dimensional case in Eq.~(\ref{eq:tilde_A_5D}), let us define a scalar field as
\be
a = \frac{1}{\sqrt{\vec D^\dagger \vec D}}\, D_a^\dagger A_a.
\ee
Then, the divergence part of Eq.~(\ref{eq:extraA_Lag_geneD}) can be written as
\be
-\dfrac{1}{2} A_a^{\rm d} \left(\delta_{ab}\square +  \xi D_aD_b^\dagger\right) A_b^{\rm d}
= -\frac{1}{2} a \left(\square + \xi \vec D^\dagger \vec D\right) a.
\ee
Since $D_a^\dagger A_a$ is orthogonal to the zero mode $d^{(0)}$ of $\vec D^\dagger \vec D$,
$a$ is also independent of $d^{(0)}$. Hence, $a$ is expanded by the eigenstates
$d^{(n)}$ of $\vec D^\dagger \vec D$ as
\be
a = \sum_{n>0}^\infty d^{(n)}(y) a^{(n)}(x).
\ee
Then contributions to the low energy effective action from the divergence part is 
\be
\int d^{D-4}y\, 
-\dfrac{1}{2} A_a^{\rm d} \left(\delta_{ab}\square +  \xi D_aD_b^\dagger\right) A_b^{\rm d}
= \sum_{n>0}^\infty
-\frac{1}{2}a^{(n)}\left(\square + \xi m_n^2\right) a^{(n)}.
\ee
This together with Eq.~(\ref{eq:expand_A4_Dgene}) for the four dimensional component $A_\mu$, 
we confirm that the massless gauge boson $A_\mu^{(0)}$ robustly exists even in higher dimensional case,
and all the KK modes $A_\mu^{(n>0)}$ become heavy by eating $a^{(n)}$.

\paragraph{The divergence-free parts}
Our last task in this subsection is clarifying mass spectra for the divergence-free
part $A_a^{\widehat{\rm df}}$, which are new degrees of freedom appearing only for $D\ge 6$. 
To this end, let us first note that the operator 
${\cal H}_{ab}$ can be expressed in the following form
\be
{\cal H}_{ab} 
= \frac{1}{(N-2)!}
\varepsilon_{i_1i_2\cdots i_{N-2}ad}
\varepsilon_{i_1i_2\cdots i_{N-2}bc}
D^\dagger_d D_c\,,
\label{eq:cal_H_1}
\ee
where $\varepsilon_{i_1\cdots i_N}$ is an $N = D-4$ dimensional completely anti-symmetric tensor.
We can rewrite this as a product of
a $\left(\begin{smallmatrix}N\\2\end{smallmatrix}\right)\times N$ matrix $\bm{D}$
and its adjoint $\bm{D}^\dagger$ as
\be
{\cal H} = \bm{D}^\dagger \bm{D}\,.
\label{eq:cal_H}
\ee
One can easily imagine the components of $\bm{D}$ from the first several examples
\be
\bm{D}\big|_{N=2} &=& \left(D_5,\,-D_4\right),
\label{eq:bmD_6}\\
\bm{D}\big|_{N=3} &=& 
\left(
\begin{smallmatrix}
0 & D_6 & -D_5\\
-D_6 & 0 & D_4\\
D_5 & -D_4 & 0
\end{smallmatrix}
\right),
\label{eq:bmD_7}\\
\bm{D}\big|_{N=4} &=&
\left(
\begin{smallmatrix}
0 & 0 & D_7 & -D_6\\
0 & -D_7 & 0 & D_5\\
0 & D_6 & -D_5 & 0\\
D_7 & 0 & 0 & -D_4\\
-D_6 & 0 & D_4 & 0\\
D_5 & -D_4 & 0 & 0
\end{smallmatrix}
\right).
\label{eq:bmD_8}
\ee
We can construct $\bm{D}\big|_{N+1}$ from $\bm{D}\big|_{N}$ as
\be
\bm{D}\big|_{N+1} = 
\left(
\begin{array}{c|cccc}
0 & & & & \\
\vdots &  &  & \bm{D}\big|_{N}(a\to a+1)   &  \\
0 & & &  & \\
\hline
(-1)^{N-1} D_{N+4} & & &  & (-1)^N D_4 \\
(-1)^{N-2} D_{N+3} & & &  (-1)^{N-1} D_4 &  \\
\vdots & &  \iddots &  & \\
D_{5} & - D_4  & &  &
\end{array}
\right),
\ee
where $\bm{D}\big|_{N}(a\to a+1)$ means $\bm{D}\big|_N$ whose indices are all shifted by 1 as
$a \to a+1$. 
Obviously, the decomposition is not unique under a unitary transformation $\bm{D} \to U\bm{D}$
with $U \in U\left(\left(\begin{smallmatrix}N\\2\end{smallmatrix}\right)\right)$.
However, this ambiguity does not yields any physical consequences. So we fix
the ambiguity by choosing a specific $\bm{D}$.
We are primally interested in existence of a massless state since its presence would be critical
in the low energy effective theory on the host topological soliton. 

Since ${\cal H}$ in Eq.~(\ref{eq:cal_H}) is semi-positive
definite, the zero eigenstate is unique and satisfies
\be
\bm{D} \vec \alpha^{(0)} = 0,\qquad
\vec\alpha^{(0)} = 
\left(
\begin{smallmatrix}
\alpha_4^{(0)}\\
\vdots\\
\alpha_{D-1}^{(0)}
\end{smallmatrix}
\right)\,.
\ee
For example, this condition in the $D=6,7,8$ cases read
\be
D=6&:&\quad D_5\alpha_4^{(0)}-D_4\alpha_5^{(0)} = 0,
\label{eq:Dalpha6}\\
D=7&:&\quad 
\left(
\begin{smallmatrix}
D_6\alpha_5^{(0)}-D_5\alpha_6^{(0)}\\
D_4\alpha_6^{(0)}-D_6\alpha_4^{(0)}\\
D_5\alpha_4^{(0)}-D_4\alpha_5^{(0)}
\end{smallmatrix}
\right) 
= 0,
\label{eq:Dalpha7}\\
D=8&:&\quad 
\left(
\begin{smallmatrix}
D_7\alpha_6^{(0)}-D_6\alpha_7^{(0)}\\
D_5\alpha_7^{(0)}-D_7\alpha_5^{(0)}\\
D_6\alpha_5^{(0)}-D_5\alpha_6^{(0)}\\
D_7\alpha_4^{(0)}-D_4\alpha_7^{(0)}\\
D_4\alpha_6^{(0)}-D_6\alpha_4^{(0)}\\
D_5\alpha_4^{(0)}-D_4\alpha_5^{(0)}
\end{smallmatrix}
\right) 
= 0.
\label{eq:Dalpha8}
\ee
These can easily be generalized in generic $D$ dimensions as
\be
D_{[a}\alpha_{b]}^{(0)} = 0,
\ee
for all $a,b = \{4,\cdots,D-1\}$.
By using the definition in Eq.~(\ref{eq:D_a}) and the assumptions $\beta_0 \neq 0$,
we can rewrite this as
\be
B_{ab} \equiv \p_a\left(\frac{\alpha_b^{(0)}}{\beta_0}\right)
-\p_b\left(\frac{\alpha_a^{(0)}}{\beta_0}\right) = 0.
\label{eq:df_massless}
\ee
Let us take any three indices from $\{4,\cdots,D-1\}$, say $4,5,6$.
For this choice, $B_{45} = B_{56} = B_{46} = 0$ is just a vorticity zero condition
to the three vector 
$\left(\frac{\alpha_4^{(0)}}{\beta_0},\frac{\alpha_5^{(0)}}{\beta_0},
\frac{\alpha_6^{(0)}}{\beta_0}\right)$.
The same is true for any choice of three indices. Therefore, 
from the conventional Stokes' theorem in three spatial dimensions,
the massless condition (\ref{eq:df_massless}) means that there exists a potential 
$-\frac{\lambda(y)}{\beta_0(y)}$ by which
any $\alpha_a^{(0)}$ can be expressed as
\be
\frac{\alpha_a^{(0)}}{\beta_0} = \p_a \left(-\frac{\lambda}{\beta_0}\right)
\quad \Leftrightarrow \quad \alpha_a^{(0)} = D_a \lambda.
\label{eq:alpha}
\ee
Finally, we have to verify if this is divergence-free or not. 
Indeed, it is a divergence part which can be seen as
\be
\int d^{D-4}y~ \vec\alpha^{(0)} \cdot \vec A^{\rm df} = 
\int d^{D-4}y~ \vec D \lambda \cdot \vec A^{\rm df} =
\int d^{D-4}y~ \lambda \vec D^\dagger \vec A^{\rm df} = 0.
\ee
Namely, the operator ${\cal H}$  has 
no divergence-free zero modes in generic $D$ dimensions.\footnote{We should mention about the result of
the previous work \cite{Arai:2018rwf} on the divergence free part. In \cite{Arai:2018rwf}, the spectrum of the
divergence free parts was studied only for $D=6$ case in detail where 
the absence of the massless mode was assumed. Furthermore, the massless modes in 
higher dimensional models ($D>6$) were not understood very well in \cite{Arai:2018rwf}.
}

In order to clarify massive modes of the divergence-free parts, 
firstly, let us introduce the eigenvectors $\bar{f}^{(n)}(y)$ and eigenvalues 
$\bar m_{\bm D,n}^2$
of the 
$\left(\begin{smallmatrix}N\\2\end{smallmatrix}\right) \times \left(\begin{smallmatrix}N\\2\end{smallmatrix}\right)$
Hermitian operator $\bar {\cal H} = \bm{D}\bm{D}^\dagger$ dual to ${\cal H}$ as
\be
\bar{\cal H} \equiv \bm{D} \bm{D}^\dagger,\qquad
\bar {\cal H} \bar{f}^{(n)} = \bar m_{\bm D,n}^2 \bar{f}^{(n)},
\ee
where $\bar{f}^{(n)}$ is a $\left(\begin{smallmatrix}N\\2\end{smallmatrix}\right)$ vector
whose component is $\bar f^{(n)}_{\tilde a}$ ($\bar a = 1,2,\cdots,
\left(\begin{smallmatrix}N\\2\end{smallmatrix}\right)$).
Then it is straightforward to show
\be
{\cal H} \bm{D}^\dagger \bar{f}^{(n)} = \bar m_{\bm D,n}^2 \bm{D}^\dagger \bar{f}^{(n)},\quad
(\bar m_{\bm D,n}\neq0).
\ee
A nice thing for this is that  the divergence-free condition is
automatically satisfied for any $\bar{f}^{(n)}$ as
\be
D^\dagger_a \left(\bm{D}^\dagger \bar{f}^{(n)}\right)_a = 0.
\ee
This can be proved by acting $D_a^\dagger$ on $(\bm{D}^\dagger)_{a\bar a}$ given
in Eqs.~(\ref{eq:Dalpha6}) -- (\ref{eq:Dalpha8}) as
\be
D_a^\dagger (\bm{D}^\dagger)_{a\bar a} = 0,\qquad
(\bm{D})_{\bar a a} D_a = 0.
\label{eq:DD}
\ee
Expanding the divergence-free parts as
\be
\vec A^{\widehat{\rm df}} = \sum_{n\neq0} L^{(n)}(x) \frac{\bm{D}^\dagger \bar{f}^{(n)}}{\bar m_{\bm D,n}},
\ee
and plugging this into the quadratic Lagrangian and integrating it over the extra dimensions,
we get
\be
\int d^{D-4}y\,\dfrac{1}{2} A^{\widehat{\rm df}}_a \Gamma^{\rm df}_{ab} A^{\widehat{\rm df}}_b 
= -\frac{1}{2}\sum_{n\neq0}L^{(n)}\left(\square + \bar m_{\bm D,n}^2\right)L^{(n)}.
\ee
This expression is formally valid for any $\beta$.

Unfortunately, it is still not clear the relation
between the eigenvalue $m_{D,n}^2$ of the $\vec D^\dagger \vec D$ operator and $\bar m_{\bm D,n}^2$ 
of the $\bar{\cal H} = \bm{D}\bm{D}^\dagger$ operator. 
For this point, the $D=6$ case is especially simple as
\be
\bar {\cal H} \big|_{D=6} = \bm{D}\bm{D}^\dagger \big|_{D=6} = D_aD_a^\dagger.
\ee

\begin{figure}[ht]
\begin{center}
\includegraphics[width=15cm]{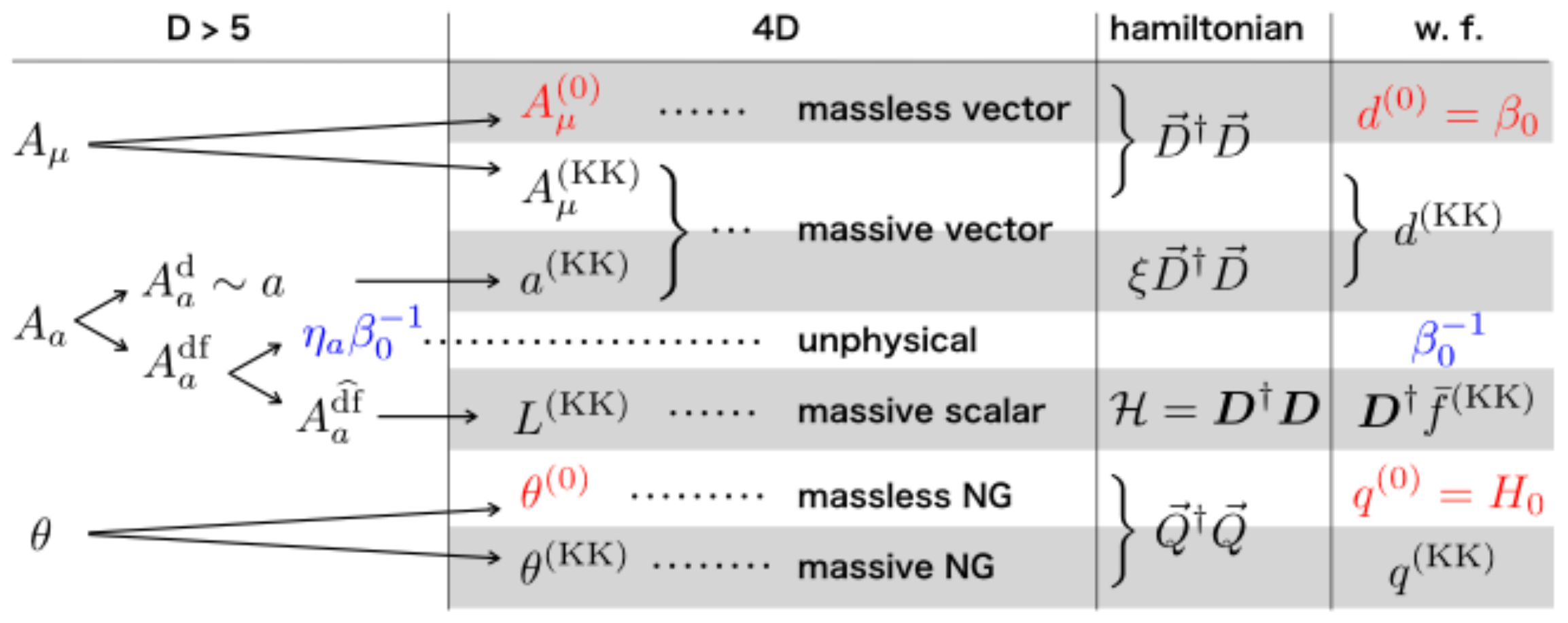}
\caption{A brief summary of the effective fields localized on 
a topological soliton by a neutral stabilizer $H$ in higher dimensional models.}
\label{fig:summary6D}
\end{center}
\end{figure}
We briefly summarize the gauge sector with Fig.~\ref{fig:summary6D}.
Similarly to the 5$D$ case,
there are only two massless states in the low energy effective theory in four dimensions:
the one is the gauge field $A_\mu^{(0)}$ and the other is the $U(1)$ NG field $\theta^{(0)}$.
In addition to the infinite KK towers of $A_\mu^{({\rm KK})}$ (eating $a^{({\rm KK})}$), $\theta^{({\rm KK})}$,
there newly appear another KK towers of $A_a^{\widehat{\rm df}({\rm KK})}$.

%

\section{Localization  by a charged stabilizer in $D=5$}
\label{sec:charged_D5}

In this section, we will again consider five dimensional models in which
the stabilizer $h$ is not neutral but has a charge to the would-be localized
$U(1)$ gauge field. Namely, we slightly modify the Lagrangian ${\cal L}_{\rm s}$ 
in Eq.~(\ref{eq:tilde_Lag}) as
\be
{\cal L}_{\rm s} = {\cal D}_MH{\cal D}^MH^* + \p_MT\p^MT - V(H,T),\quad
(M=0,1,2,3,4),
\label{eq:tilde_Lag_q}
\ee
with the conventional covariant derivative
\be
{\cal D}_M H = \p_M H + i q {\cal A}_M H.
\ee
As before, we adopt the notation that the gauge coupling constant is absorbed in the 
gauge field ${\cal A}_M$, and $q$ stands for a charge of $H$ to the $U(1)$ gauge transformation.
Except for the change $\p_MH \to {\cal D}_MH$, we will not modify the Lagrangian (\ref{eq:Lag_full}).
Thus, the Lagrangian ${\cal L}$ respects the five dimensional Lorentz symmetry and
the $U(1)$ gauge symmetry. The charged stabilizer $H$ now interacts with the $U(1)$ gauge 
field ${\cal A}_M$ through both the covariant derivative and the function $\beta(H)$
in front of the ${\cal F}^2$ term. This is a seed for a mixture of the Higgs mechanism which
takes place inhomogeneously, 
and the localization of the gauge field on the domain wall. This is what we will clarify in
this section.

The Euler-Lagrange equations are slightly modified as
\be
&&{\cal D}_M{\cal D}^M H + \frac{\p V}{\p H^*} =
-2\beta\frac{\p\beta}{\p H^*} {\cal F}_{MN}{\cal F}^{MN},\\
&&\p_M\p^M T + \frac{\p V}{\p T} = 0,\\
&&\p_M\left(\beta^2{\cal F}^{MN}\right) = iq\left(H{\cal D}^NH^* - H^* {\cal D}^N H\right).
\ee
Note that we can set $H$ to be real, and then ${\cal A}_M = 0$ solves the third equation.
The first and second equations with ${\cal A}_M = 0$ are identical with those in the 
previous section, so that the background soliton configuration with 
$T = T_0(y)$ and $H = H_0(y)$ ($H_0$ is real)
remains as the solution. In what follows, we will assume $H_0 \neq 0$.

Let us next perturb the background solution as before. The fluctuations are introduced
as $T = T_0 + \tau$, $H = e^{i\vartheta}(H_0 + h)$, and ${\cal A}_M$ itself stands
for the small fluctuation. We will obtain a quadratic Lagrangian for the fluctuations for
figuring out mass spectra. A difference from the previous model with the neutral scalar resides
only in the covariant derivative as
\be
{\cal D}_M H = e^{i\vartheta} \left\{\p_M(H_0 + h) + i H_0(\p_M \vartheta + q {\cal A}_M)\right\}.
\ee
Namely, a change from the neutral scalar case is realized by just an exchange
\be
\p_M \vartheta \to {\cal D}_M \vartheta \equiv \p_M \vartheta + q {\cal A}_M.
\ee
Therefore, the quadratic Lagrangian of the scalar sector can immediately obtained as
\be
{\cal L}_{\rm s}^{(2)} = - 
\vec\xi^{\,T}
\left(\p_\mu\p^\mu + {\cal M}^2 \right)
\vec\xi
+\, H_0^2\vartheta \left({\cal D}^\dagger_\mu {\cal D}^\mu 
- H_0^{-2} {\cal D}^\dagger_y H_0^2 {\cal D}_y\right)\vartheta,
\label{eq:tilde_L_q_D5}
\ee
where ${\cal D}_M^\dagger\vartheta = -\p_M\vartheta + q{\cal A}_M$, and 
$\vec \xi$ and ${\cal M}^2$ are defined in Eqs.~(\ref{eq:xi}) and (\ref{eq:M^2_D5}), respectively.
Let us rewrite the quadratic Lagrangian
with respect to the canonical field $\theta$ defined in Eq.~(\ref{eq:canonical_theta}) as
\be
{\cal L}_{\rm s}^{(2)} &=& - \vec\xi^{\,T} \left(\p_\mu\p^\mu + {\cal M}^2\right)
\vec\xi\nonumber\\
&&+\,  \frac{1}{2}(\p_\mu\theta + q\sqrt2 H_0{\cal A}_\mu) (\p^\mu\theta + q\sqrt2 H_0{\cal A}^\mu) \nonumber\\
&&-\, \frac{1}{2} (Q_y \theta - q\sqrt2 H_0{\cal A}_y)(Q_y\theta - q\sqrt2 H_0{\cal A}_y),
\label{eq:tilde_L_q_D5_2}
\ee
Comparing this to Eq.~(\ref{eq:tildeL_2}) for the neutral case, the terms with $q$ are added.
On the contrary, the gauge sector ${\cal L}_{\rm g}^{(2)}$ is unchanged since it is independent of $q$.
However, the scalar sector ${\cal L}_{\rm s}^{(2)}$ newly include the mixing terms
between ${\cal A}_\mu$ and $\theta$ which we would like to eliminate to diagonalize mass matrices. 
To this end, we need to modify the gauge fixing term (\ref{eq:gauge_fix}) in the five dimensions as
\be
{\cal L}_\xi = -\frac{2\beta^2}{\xi}\left\{\p_\mu {\cal A}^\mu 
- \xi\frac{1}{\beta^2}\left(
\p_y \left(\beta^2{\cal A}_y\right) + q\frac{H_0}{2\sqrt2 }\theta\right)\right\}^2.
\ee
We are now ready to write down the quadratic Lagrangian in terms of the canonically
normalized field $A_M = 2\beta_0 {\cal A}_M$ as
\be
\left({\cal L}_{\rm s} + {\cal L}_{\rm g} + {\cal L}_\xi\right)\bigg|_{\text{quad}} &=& 
-\vec\xi^{\,T}
\left(\p_\mu \p^\mu  + {\cal M}^2\right) \vec\xi\nonumber\\
&&+\,\frac{1}{2}A_\mu\left[\eta^{\mu\nu}\square - \left(1 - \frac{1}{\xi}\right)\p^\mu\p^\nu 
+ \eta^{\mu\nu}\left(
D_y^\dagger D_y + q^2{\cal E}_0^2\right)
\right] A_\nu \nonumber\\
&&-\, \frac{1}{2}A_y \square A_y - \frac{1}{2}\theta \square \theta \nonumber\\
&&-\, \frac{\xi}{2}\left(D_y^\dagger A_y + q {\cal E}_0\theta\right)^2
- \frac{1}{2}\left(Q_y \theta - q{\cal E}_0A_y\right)^2,
\label{eq:L2_5D_3}
\ee
where we defined
\be
{\cal E}_0 = \frac{H_0}{\sqrt2 \beta_0}.
\ee
Obviously, the scalar fields $\tau,h \in \vec \xi$ stand alone. Therefore, the mass spectrum 
of $\vec \xi$ is not affected by whether the stabilizer $H$ is neutral or not.
After all, all modifications from the neutral case appear only in the sectors for $A_M$ and $\theta$ 
in the specific form $q {\cal E}_0$.

The above Lagrangian includes two important phenomena. The one is the
localization of the gauge field $A_\mu$. As we saw in Eq.~(\ref{eq:L_KK_expand_5d}) in
the neutral case, the KK modes
$A_\mu^{({\rm KK})}$ get massive by absorbing $a^{({\rm KK})}$ which essentially resides in $A_y$.
The other is the conventional Higgs mechanism that a gauge field becomes massive by eating a
NG scalar field associated with a spontaneously broken symmetry. For our case, roughly speaking,
the NG is $\theta$. However, it is not precise since our soliton background is inhomogeneous.
Indeed, $A_y$ and $\theta$ are mixed in Eq.~(\ref{eq:L2_5D_3}). Our next task is to diagonalize them,
and make clear what field is physical or unphysical.

\paragraph{The simplest case}
Let us first consider the simplest example where ${\cal E}_0$ is a constant.
Then, we immediately see from Eq.~(\ref{eq:L2_5D_3})
that all the mass eigenvalues of $A_\mu$ are sifted
by a constant as
\be
m_{D,n} \to  \sqrt{m_{D,n}^2 + q^2 {\cal E}_0^2},
\ee
where $m_{D,n}$ is the eigenvalue of $D_y^\dagger D_y$ defined in Eq.~(\ref{eq:D_y^dagD_y}).
It is important to realize that now the zero eigenvalue is gone. The massless state is now lifted by 
the non zero mass $q{\cal E}_0$. This is a peculiar phenomenon which occurs as 
a consequence of interplay between the localization of gauge fields and the Higgs mechanism. 
We will explain this in more detail below.
For that purpose, let us first note that $\beta_0$ is proportional to $H_0$
when ${\cal E}_0$ is constant. This implies $Q_y = D_y$ and $Q_y^\dagger = D_y^\dagger$
from their definitions
in Eqs.~(\ref{eq:D_5}) and (\ref{eq:def_D}).
In addition, since the divergence-free part in $A_y$ is
not normalizable as is described below Eq.~(\ref{eq:5d_Lfluc_A}), we eliminate $A_y^{\rm df}$.
Hence, we can always expand $A_y = A_y^{\rm d}$ by the eigenstates $\{D_yd^{(n)}\}$ of $D_yD_y^\dagger$ 
as is given in Eq.~(\ref{eq:Ay_decompose}). At the same time, we expand $\theta$ by
the eigenstates $\{d^{(n)}\}$ of $D_y^\dagger D_y = Q_y^\dagger Q_y$ as 
\be
A_y 
=
\sum_{n\neq0}^\infty a^{(n)}(x) \frac{D_y d^{(n)}(y)}{m_{D,n}},\quad
\theta = \sum_{n=0}^\infty \theta^{(n)}(x) d^{(n)}(y).
\ee
Plugging these into the quadratic
Lagrangian (\ref{eq:L2_5D_3}) and integrating it over $y$, we find
\be
L_{\rm eff}
= L_{\rm eff}^{(0)} + \sum_{n\neq0}^\infty\left(L_{\rm eff}^{(n;1)}
+L_{\rm eff}^{(n;2)}\right),
\ee
with
\be
L_{\rm eff}^{(0)} &=& 
\frac{1}{2}A_\mu^{(0)}
\left[\eta^{\mu\nu}\square - \left(1 - \frac{1}{\xi}\right)\p^\mu\p^\nu 
+ \eta^{\mu\nu} q^2{\cal E}_0^2
\right] A_\nu^{(0)} \nonumber\\
&&-\, \frac{1}{2}\theta^{(0)}\left(\square + \xi q^2{\cal E}_0^2\right)\theta^{(0)},
\label{eq:L2_5D_4}
\ee
and
\be
L_{\rm eff}^{(n\neq0;1)} &=&
\frac{1}{2}A_\mu^{(n)}
\left[\eta^{\mu\nu}\square - \left(1 - \frac{1}{\xi}\right)\p^\mu\p^\nu 
+ \eta^{\mu\nu}\left(m_{D,n}^2+q^2{\cal E}_0^2\right)
\right] A_\nu^{(n)} \nonumber\\
&&-\, \frac{1}{2}a_q^{(n)} \left[\square + \xi \left(m_{D,n}^2+q^2{\cal E}_0^2\right)\right] a_q^{(n)},
\label{eq:L2_5D_5}\\
L_{\rm eff}^{(n\neq0;2)} &=&
- \frac{1}{2} \theta_q^{(n)} \left[\square + \left(m_{D,n}^2+q^2{\cal E}_0^2\right)\right] \theta^{(n)}_q,
\label{eq:L2_5D_6}
\ee
where we have defined new variables, in order to diagonalize the mixed terms, as
\be
a_q^{(n)} = \frac{m_{D,n} a^{(n)} + q{\cal E}_0 \theta^{(n)}}{\sqrt{m_{D,n}^2+q^2{\cal E}_0^2}},
\quad 
\theta_q^{(n)} = \frac{m_{D,n} \theta ^{(n)} - q{\cal E}_0 a^{(n)}}{\sqrt{m_{D,n}^2+q^2{\cal E}_0^2}},\quad
(n\neq0).
\ee 

The $n=0$ part (\ref{eq:L2_5D_4}) is the same form as 
a common quadratic Lagrangian of the
gauge field under the Higgs mechanism in the $R_\xi$ gauge. 
Namely, it expresses that the longitudinal mode of $A_\mu^{(0)}$ becomes physical by eating 
the Nambu-Goldstone mode $\theta^{(0)}$.
As the conventional Higgs mechanism, this occurs via the coupling $qA_\mu H$
in the covariant derivative ${\cal D}_M H$. On the other hand, regardless of the value of $q$,
the normalized physical vector fields $A_\mu^{(0)}$
appears on the domain wall thanks to the generalized gauge 
kinetic term $\beta^2{\cal F}_{MN}{\cal F}^{MN}$. This is the peculiar phenomenon 
as the consequence of interplay of the localization of the massless gauge field and
the Higgs mechanism. 

The same can be said to the massive modes $A_\mu^{(n\neq0)}$.
Indeed, Eq.~(\ref{eq:L2_5D_5}) has the same structure as Eq.~(\ref{eq:L2_5D_4}).
Instead of eating the genuine Nambu-Goldstone mode $\theta^{(n)}$, 
the KK mode $A_\mu^{(n\neq0)}$ gets massive by
absorbing $a_q^{(n\neq0)}$ which is the linear combination of 
$a^{(n)}$ and $\theta^{(n)}$. The other scalar field $\theta_q^{(n)}$ orthonormal 
to $a_q^{(n)}$ appears as a physical massive scalar field on the domain wall.

In summary, there is the unique physical light vector field $A_\mu^{(0)}$ whose mass square is
$q^2 {\cal E}_0^2$. In addition, there are the physical heavy KK vector $A_\mu^{(n\neq0)}$ 
and heavy KK scalar $\theta_q^{(n\neq0)}$ whose masses are separated from
the lightest mass $q{\cal E}_0$ by $\Omega$ of order of the inverse of the domain wall width.
Since the two mass scales independently originate, one can freely set their scales.
For example, for a phenomenological use, we may set ${\cal E}_0$ of order the electroweak scale,
whereas the domain wall scale $\Omega$ is taken to be much higher scale like GUT or the Planck scale.
In such situation, only the light massive gauge field $A_\mu^{(0)}$ is relevant in
the low energy effective theory on the domain wall \cite{Arai:2018uoy}.

\paragraph{The generic case} 
After getting an intuitive and transparent understanding through the simplest example,
let us next consider the generic case where ${\cal E}_0$ is not a constant.
We need to go back to Eq.~(\ref{eq:L2_5D_3}). The crucial difference from the simplest case
is that $Q_y$ and $D_y$ are different operators. However, they relate through
\be
{\cal E}_0 Q_y^\dagger 
= D_y^\dagger{\cal E}_0\,.
\label{eq:ID}
\ee
In order to eliminate mixed terms of $A_y$ and $\theta$  in
Eq.~(\ref{eq:L2_5D_3}), let us introduce new variables $a_q$ and $\theta_q$ by
\be
D_y^\dagger A_y 
&=& D_y^\dagger D_y 
\frac{1}{\sqrt{D_y^\dagger D_y+(q{\cal E}_0)^2}} a_q 
- \frac{1}{\sqrt{\left(D_y^\dagger D_y\right)^{-1}+(q {\cal E}_0)^{-2}}}\theta_q, \\
\theta 
&=& \frac{1}{q{\cal E}_0}\frac{1}{\sqrt{\left(D_y^\dagger D_y\right)^{-1}+(q {\cal E}_0)^{-2}}} \theta_q
+ q{\cal E}_0 \frac{1}{\sqrt{D_y^\dagger D_y+(q{\cal E}_0)^2}} a_q .
\ee
Note that $D_y^\dagger A_y$ is orthonormal to the zero mode $d^{(0)}$ of $D_y^\dagger D_y$ whereas
$\theta$ in general has non zero component for $d^{(0)}$. For consistency, 
we assume that $\theta_q$ is orthonormal to $d^{(0)}$ but
$a_q$ is not. Then, after little algebras with making use of the identity (\ref{eq:ID}), we find
\be
D_y^\dagger A_y + q{\cal E}_0 \theta 
&=& \sqrt{D_y^\dagger D_y+(q{\cal E}_0)^2}\, a_q,\\
Q_y \theta - q{\cal E}_0 A_y 
&=& {\cal K}\theta_q ,
\ee
where we have defined
\be
{\cal K} = Q_yq{\cal E}_0 \sqrt{\left(D_y^\dagger D_y\right)^{-1}+(q {\cal E}_0)^{-2}}.
\ee
With these at hand, the Lagrangian (\ref{eq:L2_5D_3}) reads
\be
{\cal L}\big|_{\text{quad}} &=& 
\frac{1}{2}A_\mu\left[\eta^{\mu\nu}\square - \left(1 - \frac{1}{\xi}\right)\p^\mu\p^\nu 
+ \eta^{\mu\nu}\left(D_y^\dagger D_y + (q{\cal E}_0)^2\right)
\right] A_\nu \nonumber\\
&-& \frac{1}{2}a_q \left[\square  + \xi \left(D_y^\dagger D_y + (q{\cal E}_0)^2\right)\right]a_q
\nonumber\\
&-& \frac{1}{2}\theta_q \left[
\square 
+ {\cal K}^\dagger {\cal K}
\right]\theta_q.
\label{eq:L2_5D_7}
\ee
We should emphasis that the mass operator for $a_q$ precisely coincides with
that of $A_\mu$ multiplied by the gauge fixing parameter $\xi$. Thus, what we need to do is expanding
$A_\nu$ and $a_q$ by the eigenfunction $d_q^{(n)}$ of the operator $D_y^\dagger D_y + (q{\cal E}_0)^2$ 
\be
\left[D_y^\dagger D_y + (q{\cal E}_0)^2\right]  d_q^{(n)} = m_{Dq,n}^2 d_q^{(n)},
\ee
as
\be
A_\mu = \sum_{n=0}^\infty A_\mu^{(n)}(x)d_q^{(n)}(y),\quad
a_q = \sum_{n=0}^\infty a_q^{(n)}(x)d_q^{(n)}(y).
\ee
Plugging these into Eq.~(\ref{eq:L2_5D_7}), we get the formally same equations 
as Eqs.~(\ref{eq:L2_5D_4}) and (\ref{eq:L2_5D_5}) with the identification $\theta^{(0)} = a_q^{(0)}$.
Interplay of the localization of gauge field and the Higgs mechanism works as follows:
the massless gauge field $A_\mu^{(0)}$ absorbs $a_q^{(0)}$ and gets non zero mass $m_{Dq,0}$.
The higher KK gauge fields $A_\mu^{(n\neq0)}$ also absorbs $a_q^{(n\neq0)}$ and becomes heavier 
than the neutral case by $m_{D,n} \to m_{Dq,n}$.
We should note that the eating and eaten fields have the exactly same wave functions $d_q^{(n)}$.

\begin{figure}[ht]
\begin{center}
\includegraphics[width=15cm]{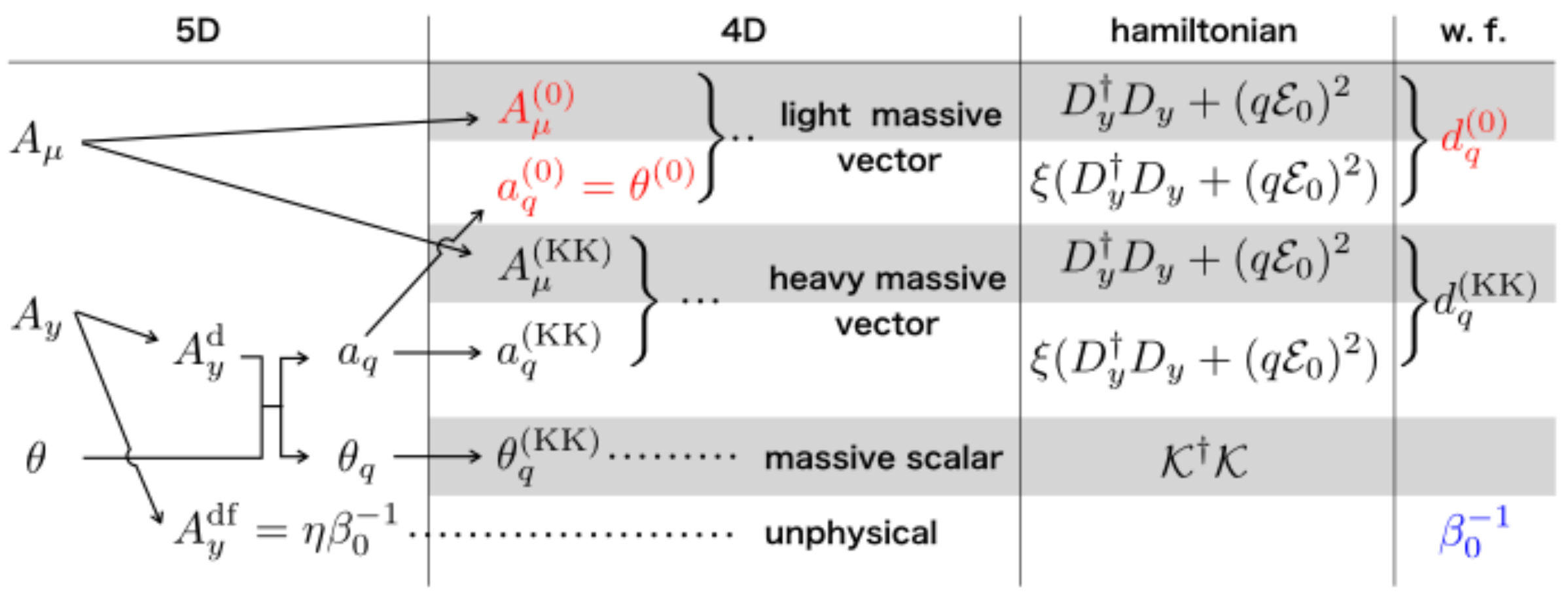}
\caption{A brief summary of the effective fields localized on 
a topological soliton by a charged stabilizer $H$ in five-dimensional models.}
\label{fig:summary5Dc}
\end{center}
\end{figure}

Fig.~\ref{fig:summary5Dc} briefly summarizes the gauge sector.
There are no massless states in the low energy effective theory in four dimensions
as a consequence of the local Higgs mechanism that
the massless gauge field $A_\mu^{(0)}$ eats the $U(1)$ NG field $\theta^{(0)}$.
In general, its mass is independent of the domain wall width, so that it is under control in the sense
that it can be light or heavy according to $\beta$. So $A_\mu^{(0)}$ is qualitatively different from
all other superheavy KK modes. 

\section{Localization by a charged stabilizer in $D\ge5$}
\label{sec:Charged_Stab_D}

We now come to analyze the most generic models with the charged stabilizer $H$. 
Namely, we will take the generic dimensions $D\ge5$, and
the generic function for $\beta(H)$. The main difference from Sec.~\ref{subsec:neutral_D}
is that the stabilizer is not neutral but has a charge $q$ to the would-be localized gauge field ${\cal A}_M$.
Furthermore, as we saw in Sec.~\ref{sec:charged_D5}, the extension from $D=5$ to $D\ge6$ is not straightforward
but we need new analysis for the divergence free parts $A_a^{\rm df}$ which would supply physical
scalar fields unlike the divergence part.

Let us begin with describing the Lagrangian again. The Lagrangian we will analyze in this section
is same as the one in Eq.~(\ref{eq:Lag_full}). We take the same gauge kinetic part as  ${\cal L}_0$ 
given in Eq.~(\ref{lagrangian_0}). However, for the scalar part, instead of ${\cal L}_{\rm s}$ in Eq.~(\ref{eq:tilde_Lag}), 
we consider ${\cal L}_{\rm s}$ given in Eq.~(\ref{eq:tilde_Lag_q}) with $M=0,1,2,\cdots,D-1$.
As before, we perturb the background configuration $T_0$ and $H_0$, and introduce small fluctuations
$\tau$, $\vartheta$, $h$, and ${\cal A}_M$ by
$T = T_0 + \tau$, $H = e^{i\vartheta}(H_0 + h)$. The quadratic Lagrangian for the scalar part
reads
\be
{\cal L}_{\rm s}^{(2)} = - 
\vec\xi^{\,T}
\left(\square + {\cal M}^2 \right)
\vec\xi
+\, H_0^2\vartheta \left({\cal D}^\dagger_\mu {\cal D}^\mu 
- H_0^{-2} {\cal D}^\dagger_a H_0^2 {\cal D}_a\right)\vartheta,
\ee
where ${\cal D}_M\vartheta = \p_M\vartheta + q{\cal A}_M$,
${\cal D}_M^\dagger\vartheta = -\p_M\vartheta + q{\cal A}_M$, and 
$\vec \xi$ and ${\cal M}^2$ are same as those given in Eq.~(\ref{eq:xi_M}).
Compared to Eq.~(\ref{eq:tilde_L_q_D5}) in $D=5$, the change is just ${\cal D}_y \to {\cal D}_a$
($a=4,5,\cdots,D-1$). Accordingly, Eq.~(\ref{eq:tilde_L_q_D5_2}) is naturally generalized as
\be
\tilde{\cal L}^{(2)} &=& - \vec\xi^{\,T} \left(\square + {\cal M}^2\right)
\vec\xi\nonumber\\
&&+\,  \frac{1}{2}(\p_\mu\theta + q\sqrt2 H_0{\cal A}_\mu) (\p^\mu\theta + q\sqrt2 H_0{\cal A}^\mu) \nonumber\\
&&-\, \frac{1}{2} (Q_a \theta - q\sqrt2 H_0{\cal A}_a)(Q_a\theta - q\sqrt2 H_0{\cal A}_a),
\ee
with the canonically normalized field $\theta = \sqrt2 H_0\vartheta$ 
defined in Eq.~(\ref{eq:canonical_theta}).
Similarly, we also extend the gauge fixing term as
\be
{\cal L}_\xi = -\frac{2\beta^2}{\xi}\left\{\p_\mu {\cal A}^\mu 
- \xi\frac{1}{\beta^2}\left(
\p_a \left(\beta^2{\cal A}_a\right) + q\frac{H_0}{2\sqrt2 }\theta\right)\right\}^2.
\ee
Correcting all the pieces ${\cal L}_{\rm s}$, ${\cal L}_{\rm g}$, and ${\cal L}_\xi$, and using
the normalized field $A_M = 2\beta_0 {\cal A}_M$, we reach the following quadratic Lagrangian
\be
{\cal L}\big|_{\text{quad}} &=& 
-\vec\xi^{\,T}
\left(\square  + {\cal M}^2\right) \vec\xi\nonumber\\
&&+\,\frac{1}{2}A_\mu\left[\eta^{\mu\nu}\square - \left(1 - \frac{1}{\xi}\right)\p^\mu\p^\nu 
+ \eta^{\mu\nu}\left(
D_a^\dagger D_a + q^2{\cal E}_0^2\right)
\right] A_\nu \nonumber\\
&&-\, \frac{1}{2}A_a \square A_a - \frac{1}{2}\theta \square \theta \nonumber\\
&&-\, \frac{\xi}{2}\left(D_a^\dagger A_a + q {\cal E}_0\theta\right)^2
- \frac{1}{2}\left(Q_a \theta - q{\cal E}_0A_a\right)^2 - \frac{1}{2}A_a {\cal H}_{ab} A_b.
\label{eq:L_quad_most_gene}
\ee
Up to here, the quadratic Lagrangian is formally trivial extension of the five dimensional case
given in Eq.~(\ref{eq:L2_5D_3}) except for the last term with ${\cal H}$
defined in Eq.~(\ref{eq:cal_H_1}). However, separating multiple fields $A_a^{\rm d}$, $A_a^{\rm df}$ 
and $\theta$ into physical and unphysical degrees of freedom is not an easy task due to the mixings
among them. 

As in the five dimensions, the quadratic Lagrangian
include $Q_a$ and $D_a$ which are related by
\be
{\cal E}_0 Q_a^\dagger 
= D_a^\dagger{\cal E}_0\,.
\label{eq:IDgene}
\ee

An idea is unifying the extra dimensional gauge fields $A_a$ and the phase $\theta$ within a $N+1$ component
vector $\vec{\mathsf{A}}$ as 
\be
\vec{\mathsf{A}} = \left(
\begin{array}{c}
\vec A\\
\theta
\end{array}
\right).
\ee
In addition, we extend $\vec D$ and ${\cal H}$ to $\vec{\mathsf{D}}$ and $\mathbb{H}$ which act on $\vec{\mathsf{A}}$ by
\be
\vec{\mathsf{D}} = 
\left(
\begin{array}{c}
\vec D\\
q{\cal E}_0
\end{array}
\right),\qquad 
\mathbb{H} = 
\left(
\begin{array}{cc}
{\cal H} + q^2{\cal E}_0^2{\bf 1}_{D-4} & -q{\cal E}_0 \vec Q\\
-q\vec Q^\dagger {\cal E}_0 & \vec Q^\dagger \vec Q
\end{array}
\right).
\ee
Similarly to Eq.~(\ref{eq:HD}), we have
\be
\mathbb{H}\, \vec{\mathsf{D}}
= 
\left(
\begin{array}{cc}
{\cal H}\vec D + q^2{\cal E}_0\left({\cal E}_0\vec D - \vec Q {\cal E}_0\right)\\
- q \vec Q^\dagger \left({\cal E}_0 \vec D - \vec Q {\cal E}_0\right)
\end{array}
\right) = 0,
\label{eq:HD2}
\ee
where we used Eqs.~(\ref{eq:HD}) and (\ref{eq:IDgene}). 
It is straightforward to show the following equations to hold
\be
\vec{\mathsf{A}}^\dagger \mathbb{H}\, \vec{\mathsf{A}} 
&=& \vec A^\dagger\left({\cal H} + q^2{\cal E}_0^2\right)\vec A
- q{\cal E}_0 \vec A^\dagger \vec Q \theta - q\theta\vec Q^\dagger {\cal E}_0 \vec A + \theta \vec Q^\dagger \vec Q \theta \nonumber\\
&=& \vec A^\dagger{\cal H} \vec A
+ \left(\vec Q \theta - q{\cal E}_0\vec A\right)^\dagger\left(\vec Q \theta - q{\cal E}_0\vec A\right),\\
\vec{\mathsf{D}}^\dagger \vec{\mathsf{A}} &=& \vec D^\dagger \vec A + q{\cal E}_0\theta,\\
\vec{\mathsf{D}}^\dagger \vec{\mathsf{D}} &=& \vec D^\dagger \vec D + q^2{\cal E}_0^2.
\ee
Then, we can rewrite the quadratic Lagrangian (\ref{eq:L_quad_most_gene}) into the following 
compact form as
\be
{\cal L}\big|_{\text{quad}} &=& 
-\vec\xi^{\,T}
\left(\square  + {\cal M}^2\right) \vec\xi\nonumber\\
&&+\,\frac{1}{2}A_\mu\left[\eta^{\mu\nu}\square - \left(1 - \frac{1}{\xi}\right)\p^\mu\p^\nu 
+ \eta^{\mu\nu}\vec{\mathsf{D}}^\dagger \vec{\mathsf{D}}
\right] A_\nu \nonumber\\
&&-\, \frac{1}{2}\vec{\mathsf{A}}^\dagger \left(\square + \mathbb{H} + \xi \vec{\mathsf{D}}\vec{\mathsf{D}}^\dagger
\right)\vec{\mathsf{A}}.
\label{eq:L_quad_most_gene_2}
\ee
Note that the gauge sector (the second and third lines) formally coincide with 
the quadratic Lagrangian of the gauge part in the neutral case
Eq.~(\ref{lagrangian_0_4ex}) with $\Gamma$s given in Eqs.~(\ref{eq:Gamma4}) and (\ref{eq:Gamma5}).
Therefore, we can repeat the almost same procedures to decompose physical and unphysical
degrees of freedom from $\vec{\mathsf{A}}$. Firstly, let us define a projection operator
\be
\mathbb{P} = \vec{\mathsf{D}}\left(\vec{\mathsf{D}}^\dagger\vec{\mathsf{D}}\right)^{-1}\vec{\mathsf{D}}^\dagger.
\ee 
Then we decompose $\vec{\mathsf{A}}$ as
\be
\vec{\mathsf{A}} = \vec{\mathsf{A}}^{\rm d} + \vec{\mathsf{A}}^{\rm df},\quad
\vec{\mathsf{A}}^{\rm d} \equiv \mathbb{P}\vec{\mathsf{A}},\quad
\vec{\mathsf{A}}^{\rm df} \equiv  \left(\mathsf{1}-\mathbb{P}\right)\vec{\mathsf{A}}.
\label{eq:d_df_D}
\ee
Since the projection operator includes $\left(\vec{\mathsf{D}}^\dagger\vec{\mathsf{D}}\right)^{-1}$,
we should check if it is always regular or not. It would be ill-defined when
$\left(\vec{\mathsf{D}}^\dagger\vec{\mathsf{D}}\right)^{-1}$ acts on a zero mode  of
$\vec{\mathsf{D}}^\dagger\vec{\mathsf{D}}$.
However, it is obvious that $\vec{\mathsf{D}}^\dagger\vec{\mathsf{D}} = \vec D^\dagger \vec D + q^2{\cal E}_0^2$
does not have zero eigenstates unless $q{\cal E}_0 = 0$. Thus, we proved that the
projection operator $\mathsf{P}$ is well-defined as long as $q\neq0$.

From Eqs.~(\ref{eq:HD2}) and (\ref{eq:d_df_D}), we find
\be
\mathbb{H}\vec{\mathsf{A}}^{\rm d} = 0,\quad
\vec{\mathsf{D}}^\dagger \vec{\mathsf{A}}^{\rm df} = 0.
\ee
The divergence free part can be decomposed as
\be
\vec{\mathsf{A}}^{\rm df} =
\left(
\begin{array}{c}
\beta_0^{-1}\vec\eta\\
0
\end{array}
\right) + 
\vec{\mathsf{A}}^{\widehat{\rm df}}.
\ee
The first term in the right hand side diverges at the spatial infinity, so we remove it by hand.
With these at hand, now, we can rewrite
\be
\vec{\mathsf{A}}^\dagger \left(\square + \mathbb{H} + \xi \vec{\mathsf{D}}\vec{\mathsf{D}}^\dagger
\right)\vec{\mathsf{A}}
&=& \vec{\mathsf{A}}^{\widehat{\rm df}\dagger}\left(\square + \mathbb{H}\right) \vec{\mathsf{A}}^{\widehat{\rm df}} + 
\vec{\mathsf{A}}^{{\rm d}\dagger}\left(\square 
+ \xi \vec{\mathsf{D}}\vec{\mathsf{D}}^\dagger\right) \vec{\mathsf{A}}^{{\rm d}} \nonumber\\
&=& \vec{\mathsf{A}}^{\widehat{\rm df}\dagger}\left(\square + \mathbb{H}\right) \vec{\mathsf{A}}^{\widehat{\rm df}} + 
\mathsf{a}^\dagger
\left(\square 
+ \xi \vec{\mathsf{D}}^\dagger\vec{\mathsf{D}} \right)
\mathsf{a},
\ee
where we defined
\be
\mathsf{a} \equiv 
\frac{1}{\sqrt{\vec{\mathsf{D}}^\dagger\vec{\mathsf{D}}}}\vec{\mathsf{D}}^\dagger \vec{\mathsf{A}}.
\ee
Thus, we again run into the coincidence between the mass determining operator $\vec{\mathsf{D}}^\dagger\vec{\mathsf{D}}$
for $A_\mu$ and $\xi\vec{\mathsf{D}}^\dagger\vec{\mathsf{D}}$ for $\mathsf{a}$. This implies that
$\mathsf{a}$ is eaten by $A_\mu$ to give non zero masses.
For the divergence free part, we mimic the factorization done in Eq.~(\ref{eq:cal_H}).
In this case, $\mathbb{H}$ can be factorized as
\be
\mathbb{H} = \mathfrak{D}^\dagger \mathfrak{D},
\ee
with
\be
\mathfrak{D} = \left(
\begin{array}{cc}
\bm{D} & 0\\
q{\cal E}_0{\bf 1}_{N} & - \vec Q
\end{array}
\right),
\ee
where $\bm{D}$'s are given in Eqs.(\ref{eq:bmD_6}) -- (\ref{eq:bmD_8}).
The size of $\mathfrak{D}$ is $\left(\begin{smallmatrix}N+1\\2\end{smallmatrix}\right)
\times (N+1)$.
Let us study a zero mode of $\mathbb{H}$ which satisfies
\be
\mathbb{H}\,\vec{\mathsf{f}}^{\, (0)} = 0
\quad \Leftrightarrow \quad
\mathfrak{D} \vec{\mathsf{f}}^{\, (0)} = 0
\quad \Leftrightarrow \quad
\vec{\mathsf{f}}^{\, (0)} = \vec{\mathsf{D}} g,
\ee
with an arbitrary scalar function $g$. One can directly verify this as
\be
\mathfrak{D}\vec{\mathsf{D}} =  
\left(
\begin{array}{cc}
\bm{D} & 0\\
q{\cal E}_0{\bf 1}_{N} & - \vec Q
\end{array}
\right)
\left(
\begin{array}{c}
\vec D\\
q{\cal E}_0
\end{array}
\right)
=
\left(
\begin{array}{c}
\bm{D}\vec D\\
q{\cal E}_0\vec D - \vec Q(q{\cal E}_0)
\end{array}
\right) = 0,
\ee
where we used Eqs.~(\ref{eq:DD}) and (\ref{eq:IDgene}).
This implies that the divergence-free part is orthogonal to $\vec{\mathsf{f}}^{\, (0)}$ because
\be
\int d^{D-4}y\, \vec{\mathsf{f}}^{\, (0)}\cdot\vec{\mathsf{A}}^{\rm df} =
\int d^{D-4}y\, \vec{\mathsf{D}}g\cdot\vec{\mathsf{A}}^{\rm df} =
\int d^{D-4}y\, g\vec{\mathsf{D}}^\dagger\vec{\mathsf{A}}^{\rm df} = 0.
\ee

Now, we consider a dual operator 
\be
\bar{\mathbb{H}} = \mathfrak{D}\mathfrak{D}^\dagger,\qquad
\bar{\mathbb{H}}\, \bar{\mathsf{f}}^{(n)} = \bar m_{\mathfrak{D},n}^2 \bar{\mathsf{f}}^{(n)},
\ee
where $\bar{\mathsf{f}}^{(n)}$ is a $\left(\begin{smallmatrix}N+1\\2\end{smallmatrix}\right)$ vector.
Then one can easily verify the following equations
\be
\mathbb{H} \left(\mathfrak{D}^\dagger \bar{\mathsf{f}}^{(n)} \right)
= \bar m_{\mathfrak{D},n}^2 \left(\mathfrak{D}^\dagger \bar{\mathsf{f}}^{(n)} \right),
\qquad
\vec{\mathsf{D}}^\dagger \left(\mathfrak{D}^\dagger \bar{\mathsf{f}}^{(n)}\right) = 0.
\label{eq:HDF}
\ee
Thus, the divergence-free component $\vec{\mathsf{A}}^{\widehat{\rm df}}$ can be decomposed by
the divergence-free eigenfunctions $\mathfrak{D}^\dagger \bar{\mathsf{f}}^{(n)}$ of $\mathbb{H}$ as
\be
\vec{\mathsf{A}}^{\widehat{\rm df}} = \sum_n \mathfrak{L}^{(n)}(x) \frac{\mathfrak{D}^\dagger \bar{\mathsf{f}}^{(n)}}{\bar m_{\mathfrak{D},n}},
\ee
and the divergence-free part reads
\be
\int d^{D-4}y\,\vec{\mathsf{A}}^{\widehat{\rm df}\dagger}\left(\square + \mathbb{H}\right) \vec{\mathsf{A}}^{\widehat{\rm df}}
= \sum_n \mathfrak{L}^{(n)}\left(\square + \bar m_{\mathfrak{D},n}^2\right)\mathfrak{L}^{(n)}.
\ee
Hence, the formulae for the charged stabilizer are the same as those for the neutral stabilizer
if we replace $\left(\vec{\mathsf{A}}^{\rm d},\vec{\mathsf{A}}^{\widehat{\rm df}},\vec{\mathsf{D}},\mathfrak{D},\mathbb{H},
\bar{\mathbb{H}}\right)$ by
$\left(\vec{A}^{\rm d},\vec{A}^{\widehat{\rm df}},\vec{D},\bm{D},{\cal H},\bar{\cal H}\right)$.
%
%


Let us compare the results in 5 dimensions given in Sec.~\ref{sec:charged_D5} and
in higher dimensions obtained in this section.
In the five dimensional case, we decomposed the physical ($\theta_q$) and unphysical ($a_q$)
degrees of freedom as given in Eq.~(\ref{eq:L2_5D_7}). However, the mass square operator 
${\cal K}^\dagger {\cal K}$ is the complicated  operator, so that it is difficult to
obtain eigenvalues in reality. Compared to this, 
the generic formula given here is better since 
obtaining mass eigenvalues of $\vec{\mathsf{A}}^{\widehat{\rm df}}$ 
is relatively easier. This is because its mass square operator $\mathbb{H}$ remains simple
as a consequence of unifying $\vec A$ and $\theta$ in $\vec{\mathsf{A}}$.

\begin{figure}[ht]
\begin{center}
\includegraphics[width=15cm]{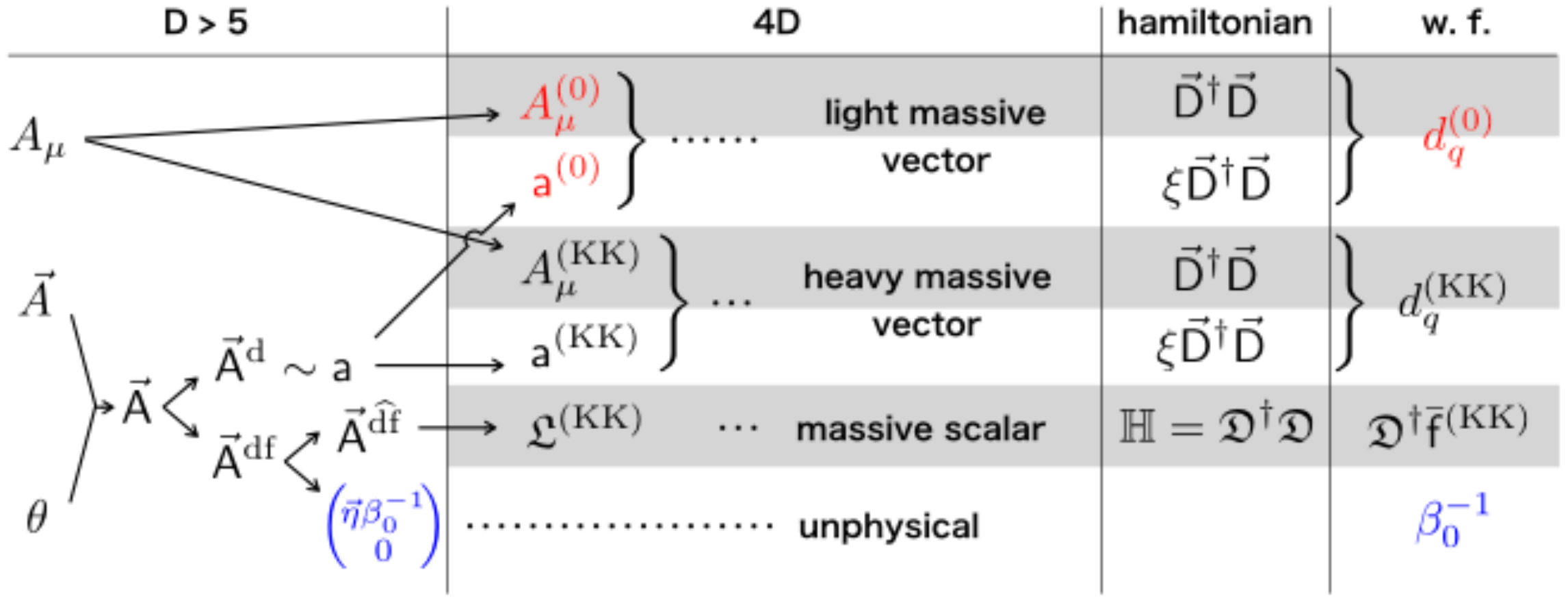}
\caption{A brief summary of the effective fields localized on 
a topological soliton by a charged stabilizer $H$ in higher dimensional models.}
\label{fig:summary6Dc}
\end{center}
\end{figure}

Fig.~\ref{fig:summary6Dc} briefly summarizes the gauge sector.
As in the 5$D$ case, there are no massless states in the low energy effective theory in four dimensions.
The lightest field is the massive vector field $A^{(0)}_\mu$.
All other physical degrees of freedom reside in $A_\mu^{({\rm KK})}$ and the divergence-free components
$\vec{\mathsf{A}}^{\widehat{\rm df}({\rm KK})}$. They are superheavy whose masses are of order inverse
of the domain wall width.

Before closing this section, let us make a comment on the divergence free part for a model
with a constant ${\cal E}_0=H_0/(\sqrt{2}\beta_0)$.  From the condition 
$\vec{\mathsf{D}}^\dagger \vec{\mathsf{A}}^{\rm df} = 0$, we find
\be
\theta = - \frac{1}{q{\cal E}_0}\, \vec D^\dagger \vec A
= - \frac{1}{q{\cal E}_0}\vec D^\dagger \left(\vec A^{\rm d} + \vec A^{\rm df}\right)
= - \frac{1}{q{\cal E}_0}\vec D^\dagger \vec A^{\rm d}.
\ee
Therefore, we can express the divergence free part as
\be
\vec{\mathsf{A}}^{\rm df} = 
\left(
\begin{array}{c}
\vec A^{\rm df}\\
0
\end{array}
\right) + 
\left(
\begin{array}{c}
\vec A^{\rm d}\\
-\frac{1}{q{\cal E}_0}\vec D^\dagger \vec A^{\rm d}
\end{array}
\right) = 
\left(
\begin{array}{c}
\bm{D}^\dagger\\
0
\end{array}
\right)\bar{f} + 
\left(
\begin{array}{c}
1\\
-\frac{1}{q{\cal E}_0}\vec D^\dagger
\end{array}
\right) \vec A^{\rm d},
\label{eq:Adf_sf}
\ee
where we used $\vec A^{\rm df} = \bm{D}^\dagger \bar{f}$ from Eq.~(\ref{eq:DD}).
Note that the first and the second terms in the right hand side are orthogonal each other.
This can be easily verified as
\be
\left(
\begin{array}{c}
\bm{D}^\dagger\\
0
\end{array}
\right)^\dagger
\left(
\begin{array}{c}
\vec D(\vec D^\dagger \vec D)^{-1}\\
-\frac{1}{q{\cal E}_0}
\end{array}
\right)
= \bm{D}\vec D(\vec D^\dagger \vec D)^{-1} = 0,
\ee
where we used $\bm{D}\vec D = 0$ from Eq.~(\ref{eq:DD}).
Since ${\cal E}_0$ is constant, $\vec D = \vec Q$. Then we have
\be
\mathbb{H} \left(
\begin{array}{c}
\bm{D}^\dagger\\
0
\end{array}
\right) 
&=& \left(
\begin{array}{c}
\bm{D}^\dagger\\
0
\end{array}
\right)\left(\bar{\cal H} + q^2{\cal E}_0^2{\bf 1}_{\left(\begin{smallmatrix}N\\2\end{smallmatrix}\right)}\right),\\
\mathbb{H}
\left(
\begin{array}{c}
1\\
-\frac{1}{q{\cal E}_0}\vec D^\dagger
\end{array}
\right)
&=& \left(
\begin{array}{c}
1\\
-\frac{1}{q{\cal E}_0}\vec D^\dagger
\end{array}
\right) \left(\vec D \vec D^\dagger + q^2{\cal E}_0^2 {\bf 1}_N\right),
\ee
where we used Eq.~(\ref{eq:DD}) for $\vec D^\dagger \bm{D}^\dagger = 0$
and Eq.~(\ref{eq:id1}) for ${\cal H} \vec A^{\rm d} = 0$.
Hence, to see the physical spectra in $\vec{\mathsf{A}}^{\rm df}$, we need to expand $\bar{f}$
by the eigenstates of $\bar{\cal H}$ operator, and expand $\vec A^{\rm d}$ by
the eigenstates of $\vec D \vec D^\dagger$ operator.

\section{Several examples in $D=6$}
\label{sec:example}

\subsection{Intersection of domain walls}


Let us consider the scalar Lagrangian in six dimensions
\be
{\cal L}_{{\rm s}} = |\p_M H|^2 - \Omega^2|H|^2 
+ \sum_{i=4,5}\left[(\p_M T_i)^2 - \lambda^2\left(T_i^2 + |H|^2 - v^2\right)^2\right].
\label{eq:L_s_intersection}
\ee
Here, $T_4$ and $T_5$ are real scalar fields, and $H$ is a complex scalar field.
There exist four discrete vacua $T_4 = \pm v$ and $T_5 = \pm v$ with $H = 0$.
Thus, we have two kinds of domain walls associated with the discrete symmetry
$\mathbb{Z}_2 \times \mathbb{Z}_2$: the one made of $T_4$ and the other 
made of $T_5$. 
The domain walls, in general, are not parallel each other, and intersect at an angle.
Hereafter, we will concentrate on the intersecting domain walls at 90 degree,
see Refs.~\cite{Gauntlett:2000bd,Gauntlett:2000ib,Eto:2005sw,Eto:2006pg} for
the intersecting domain walls in supersymmetric models.  

\begin{figure}[ht]
\begin{center}
\includegraphics[width=15cm]{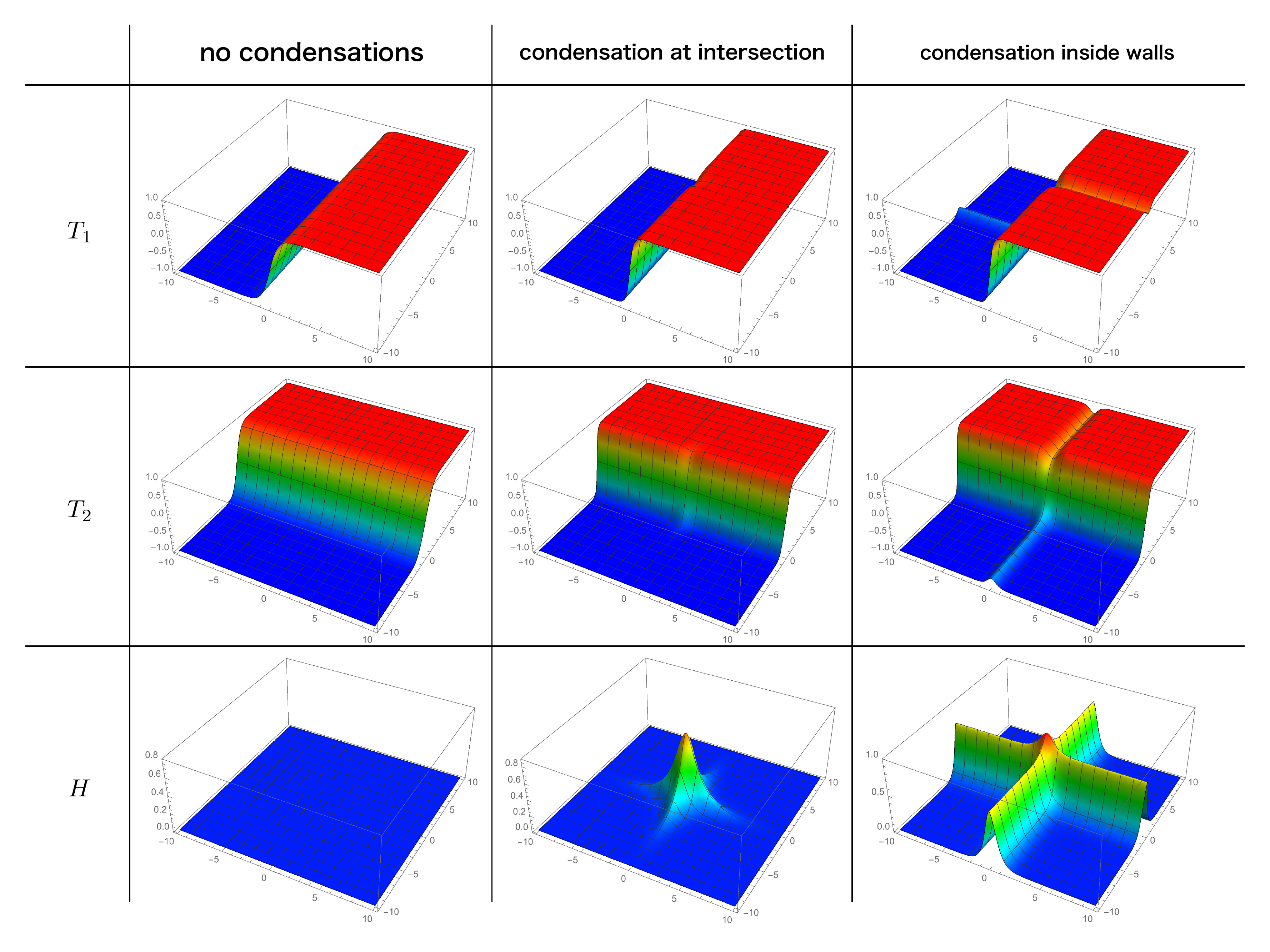}
\caption{Three typically different numerical solutions. $T_{1,2}$ and $H$ are plotted on the $x^4$-$x^5$
plane. As a benchmark, we take $\Omega = 2$ and $\lambda = 1$. We also take
$v=1.1$, $1.9$, and $2.5$ for the left-, middle-, and right-column, respectively.}
\label{fig:intersection}
\end{center}
\end{figure}
Similarly to the single domain wall studied in Sec.~\ref{sec:5D_simplest}, each domain wall
can induce local condensation of $H$ according to values of the parameters.
To quickly see this, let us make an ansatz
\be
T_4 = v \tanh \Omega x^4,\quad
T_5 = v \tanh \Omega x^5,\quad H = 0.
\ee
We fix $T_4$ and $T_5$ by hand, and perturb $H$ by $H = 0 + h$ (for simplicity, we
will omit the $U(1)$ NG mode and set $h$ to be real). Then, we find the Schr\"odinger potential for $h$ as
\be
V_h = \Omega^2  - 2\lambda^2v^2 \left({\rm sech}^2\Omega x^4 + {\rm sech}^2\Omega x^5\right).
\ee
Depth of the potential is two folds: $V_h \to \Omega^2 - 2\lambda^2 v^2$ inside the each wall,
and $V_h \to \Omega^2 - 4\lambda^2v^2$ at the wall intersection. Thus, we expect that a tachyonic mode
of $h$ appears, and it is localized either on the domain walls or only at the intersecting point
according to $\lambda v/ \Omega$.
In order verify this observation, we numerically solved equations of motion for 
the model (\ref{eq:L_s_intersection}).
The numerical solutions are shown in Fig.~\ref{fig:intersection}.
We found three qualitatively different configurations. The first solution (the left column of
Fig.~\ref{fig:intersection}) has no $H$ condensations at all.
This appears when $\Omega$ is sufficiently larger than $\lambda v$.
The second solution has finite $H$ condensation only around the intersecting point
(the middle column of Fig.~\ref{fig:intersection}).
The third solution has infinite $H$ condensation along the domain walls
(the right column of Fig.~\ref{fig:intersection}).
For our purpose of constructing four dimensional low energy theory, we prefer the finite
condensation of $H$ whose codimension is two in six dimensions. Therefore, we will consider
solutions of the type of middle column of Fig.~\ref{fig:intersection}.

\begin{figure}[th]
\begin{center}
\includegraphics[width=13cm]{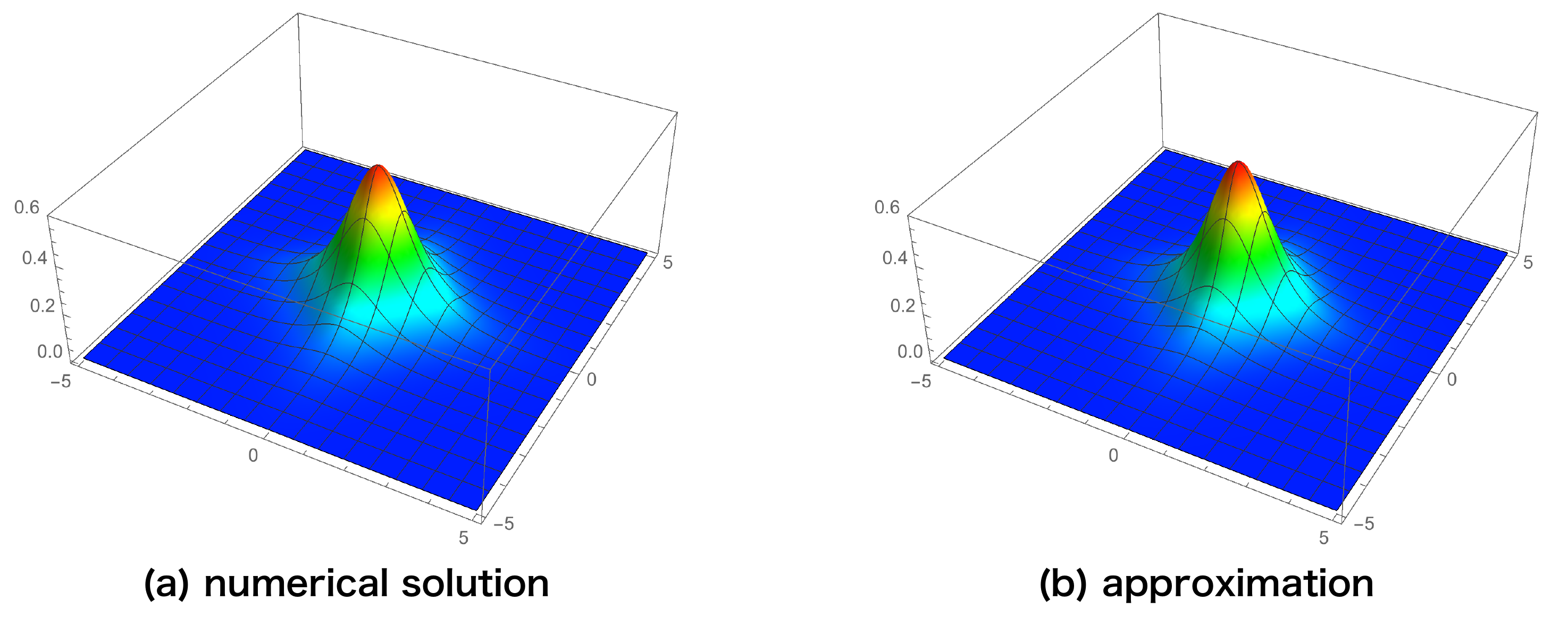}
\caption{The left panel shows a numerical solution for $(\Omega,\lambda,v) = (2,1,1.5)$.
The right panel is a plot of the separated approximation with $(a,\Omega')=(0.6,\sqrt2)$ in
Eq.~(\ref{eq:H_approx}).}
\label{fig:approximation}
\end{center}
\end{figure}

Unfortunately, we do not have an analytic solution of the intersecting domain walls with
a finite and non-zero condensation of $H$. However, for an appropriate
choice of parameters $\lambda$, $v$ and $\Omega$,
we can make a simple separated approximation
\be
H_{\rm approx} = a\, {\rm sech}\,\Omega' x^4~ {\rm sech}\,\Omega' x^5,
\label{eq:H_approx}
\ee
with $a$ and $\Omega'$ being approximation parameters.
We assume $\Omega'$ is the same order of $\Omega$.
This product ansatz works very well as can be seen in Fig.~\ref{fig:approximation} where we compare
the numerical solution and the approximation with appropriate $a$ and $\Omega'$.

With the above separable background condensation in Eq.~(\ref{eq:H_approx}) at hand, 
we will now investigate localization 
of the gauge field on the intersecting point through the Lagrangian
\be
{\cal L}_{\rm g} = - \beta^2 {\cal F}_{MN}{\cal F}^{MN},\qquad (M,N=0,1,2,3,4,5),
\ee
with
\be
\beta = \frac{|H|}{2\mu}\quad \to \quad \beta_0 \simeq \frac{a}{2\mu}
{\rm sech}\,\Omega' x^4~ {\rm sech}\,\Omega' x^5\,.
\label{eq:beta_D6}
\ee
The separable property of the approximate function $\beta_0$ in $x^4$ and $x^5$ will
help us to obtain the mass spectra below.

\subsubsection{Neutral stablizer}

Here, we study the mass spectra of $\theta$, $A_\mu$, and $A_a$ for the model with
the neutral stabilizer $H$.
All the formulae are given in Sec.~\ref{subsec:neutral_D}.
As is given in Eq.~(\ref{eq:Lag_2s_D}), the mass spectrum for $\theta$ is determined by
$Q_a^\dagger Q_a$. Similarly, the mass spectrum of $A_\mu$ is determined by $D_a^\dagger D_a$
from Eq.~(\ref{eq:Lag_2Amu_D}). 
In general, $Q_a$ and $D_a$ are different, but for the special choice of $\beta$
proportional to $|H|$ in Eq.~(\ref{eq:beta_D6}) they are identical. To obtain their mass spectra,
we will make use of the separable approximation for $\beta_0$ in Eq.~(\ref{eq:beta_D6}).
In this approximation, we have
\be
D_a \simeq - \p_a + \frac{\p_a \,{\rm sech}\,\Omega' x^a}{{\rm sech}\,\Omega' x^a}
\quad \to \quad
D_a^\dagger D_a \simeq \sum_{a=4,5}\left[- \p_a^2 + \Omega'{}^2\left(1-2\,{\rm sech}\,^2 \Omega' x^a\right)\right].
\ee
Hence, the eigenvalue equation $D_a^\dagger D_a d_n = m_{D,n}^2 d_n$ can be solved by 
separation of variables about $x^4$ and $x^5$. The separated equations are identical to
Eq.~(\ref{eq:DD_5d}) which we have already analytically solved. There exists the unique zero
eigenstate 
\be
m_{D,0} = 0,\quad d^{(0)} \propto\, {\rm sech}\, \Omega' x^4\, {\rm sech}\, \Omega' x^5.
\ee
All the excited states are continuum scattering modes as Eq.~(\ref{eq:continuum_sech}) whose
masses are given by 
\be
m_D(k_4,k_5)^2 = \left\{k_4^2 + \Omega'{}^2,\ k_5^2 + \Omega'{}^2,\ k_4^2 + k_5^2 + 2\Omega'{}^2\right\}.
\ee

The mass spectrum of the divergence-free part 
$A_a^{\widehat{\rm df}}$ is determined by ${\cal H}$ from Eq.~(\ref{eq:extraA_Lag_geneD}),
with
\be
\bm{D} = \left(
\begin{array}{cc}
D_5 & - D_4
\end{array}
\right),\qquad
{\cal H} = \bm{D}^\dagger \bm{D} =
\left(
\begin{array}{cc}
D_5^\dagger D_5 & - D_5^\dagger D_4 \\
-D_4^\dagger D_5 & D_4^\dagger D_4
\end{array}
\right).
\ee
We have shown in Sec.~\ref{sec:A_a_neutral_D} 
that the mass spectrum of ${\cal H}$ is identical to that of $\bar{\cal H}$ which can
again be solved by separation of variables in $x^4$ and $x^5$. For the specific $\beta_0$ 
given in Eq.~(\ref{eq:beta_D6}), the problem becomes even simpler since $\bar{\cal H}$ has a constant
potential:
\be
\bar{\cal H} = {\bm D}{\bm D}^\dagger = D_aD_a^\dagger
= - \p_a^2 + 2 \Omega'{}^2.
\ee
Therefore, there are no bound states and the mass spectrum is given by
\be
\bar m_{\bm{D},n}^2 = k_4^2 + k_5^2 + 2\Omega'{}^2.
\label{eq:div_free_mass_neutral}
\ee

Thus, we conclude that the massless bound states localized at the intersection point are
the four dimensional gauge field $A_\mu^{(0)}$ and the Nambu-Goldstone field $\theta^{(0)}$.
All other KK states from $A_M$ and $\theta$
are superheavy with the mass of order $\Omega'$ and are not localized.

\subsubsection{Charged stabilizer}

We next study the same model as Eq.~(\ref{eq:L_s_intersection}), but now the stabilizer $H$
is not neutral. So, we replace $\p_M H$ by ${\cal D}_M H$.
The mass spectrum of $A_\mu$ in this case is determined by 
$\vec{\mathsf{D}}^\dagger \vec{\mathsf{D}} = D_a^\dagger D_a + q^2{\cal E}_0^2$, see 
Eq.~(\ref{eq:L_quad_most_gene_2}). For the choice of $\beta$ in Eq.~(\ref{eq:beta_D6}),
${\cal E}_0$ is a constant
\be
{\cal E}_0 = \frac{H_0}{\sqrt 2\, \beta_0} = \sqrt2\,\mu.
\ee
Thus, the masses of $A_\mu^{(n)}$ are all shifted by the constant $\sqrt2\,q\mu$ from
those of the neutral case. Namely, there is only one bound state $A_\mu^{(0)}$ which
gets mass $\sqrt2\,q\mu$ by the Higgs mechanism which locally occurs at the intersection point.

The remaining fields $\theta$ and $A_a$ are unified in $\vec{\mathsf{A}}$, and the physical degrees
of freedom are confined in $\vec{\mathsf{A}}^{\widehat{\rm df}}$. The mass spectrum of 
$\vec{\mathsf{A}}^{\widehat{\rm df}}$ are
identical to the eigenvalues of $\mathbb{H}$:
\be
\mathfrak{D} &=& \left(
\begin{array}{ccc}
D_5 & - D_4 & 0\\
q{\cal E}_0 & 0 & - D_4\\
0 & q{\cal E}_0 & - D_5
\end{array}
\right),\\
\mathbb{H} &=& \mathfrak{D}^\dagger \mathfrak{D} = 
\left(
\begin{array}{ccc}
D_5^\dagger D_5 + q^2{\cal E}_0^2 & - D_5^\dagger D_4 & -q{\cal E}_0D_4\\
-D_4^\dagger D_5 & D_4^\dagger D_4 + q^2{\cal E}_0^2 & - q{\cal E}_0D_5\\
- q {\cal E}_0 D_4^\dagger  & - q{\cal E}_0 D_5^\dagger & D_4^\dagger D_4 + D_5^\dagger D_5
\end{array}
\right).
\ee
Following the general arguments of Sec.~\ref{sec:Charged_Stab_D}, the eigenvalues of $\mathbb{H}$
coincides with $\bar{\mathbb{H}}$ whose explicit form is given by
\be
\bar{\mathbb{H}} = \mathfrak{D}\mathfrak{D}^\dagger = 
\left(
\begin{array}{ccc}
D_4D_4^\dagger + D_5D_5^\dagger & q {\cal E}_0D_5 & - q {\cal E}_0D_4\\
q{\cal E}_0D_5^\dagger & D_4D_4^\dagger + q^2{\cal E}_0^2 & D_4 D_5^\dagger\\
-q{\cal E}_0 D_4^\dagger & D_5 D_4^\dagger & D_5D_5^\dagger + q^2{\cal E}_0^2
\end{array}
\right).
\ee
Hence, in contrast to ${\cal H}$ and $\bar {\cal H}$ in the neutral case,
there are no practical advantages to deal with $\bar{\mathbb{H}}$ instead of $\mathbb{H}$.  

In order to obtain the mass spectra of the divergence free part, we make use of the generic
argument given at the end of the Sec.~\ref{sec:Charged_Stab_D}. According to them, the 
divergence free part can be decomposed into two orthogonal components as Eq.~(\ref{eq:Adf_sf}).
The mass spectrum for the component associated with $\bar{f}$ of (\ref{eq:Adf_sf}) 
corresponds to the eigenvalues of $\bar {\cal H} + 2 q^2\mu^2$, and the one for the other component 
$\vec A^{\rm d}$ 
corresponds to the eigenvalues of $\vec D \vec D^\dagger + 2q^2\mu^2{\bf 1}_2$. 
In the six dimensions, we have $\bar{\cal H} = D_a D_a^\dagger$ and
$\vec D\vec D^\dagger = \left(\begin{smallmatrix}D_4D_4^\dagger & D_4 D_5^\dagger\\ D_5 D_4^\dagger & D_5D_5^\dagger\end{smallmatrix}\right)$.
Note that the non-zero eigenvalues of $D_aD_a^\dagger$ are identical to those of $\vec D^\dagger \vec D = D_a^\dagger D_a$,
if $\beta_0$ is separable. On the other hand, the non-zero eigenvalues of 
$\vec D \vec D^\dagger$ and $\vec D^\dagger \vec D$ are always identical.
Therefore, for the separable $\beta_0$, the two orthogonal components in the 
divergence free part $\vec{\mathsf{A}}^{\rm df}$ (the first and the second term of Eq.~(\ref{eq:Adf_sf}))
are degenerate with eigenvalues  
of $D_a D_a^\dagger + 2q^2\mu^2 = - \p_a^2 + 2\Omega'{}^2 + 2q^2\mu^2$. Therefore, no light bound states exists
and all the massive modes are heavy scattering modes whose masses are of order ${\cal O}(\Omega')
\sim {\cal O}(\Omega)$.

\subsection{Axially symmetric case}

Our next example has $\beta_0$ that is not separable  but is axially symmetric in the $x^4$-$x^5$ plane.
To be concrete, we assume the following Gaussian 
\be
H = a e^{-\Omega^2 r^2},\quad r^2 = (x^4)^2 + (x^5)^2.
\ee
Furthermore, we consider a specific $\beta$ as before
\be
\beta = \frac{|H|}{2\mu} \quad \to \quad
\beta_0 = \frac{a}{2\mu}e^{-\Omega^2 r^2}.
\label{eq:beta_D6_2}
\ee

\subsubsection{Neutral stablizer}

As before, we concentrate on $\theta$, $A_\mu$, and $A_a$.
All the formulae are given in Sec.~\ref{subsec:neutral_D}.
For our special choice of $\beta$
proportional to $|H|$ in Eq.~(\ref{eq:beta_D6_2}),
$\theta$ and $A_\mu$ have the same mass
eigenvalues of
\be
D_a = - \p_a + \frac{\p_a e^{-\Omega^2 r^2}}{e^{-\Omega^2 r^2}}
\quad \to \quad
D_a^\dagger D_a = - \p_r^2 - \frac{1}{r}\p_r - \frac{1}{r^2}\p_\phi^2 + 4\Omega^2\left(\Omega^2 r^2 - 1\right),
\ee
with $x^4+i x^5 = r e^{i\phi}$. The Schr\"odinger potential is asymptotically $4\Omega^4r^2$, so all the eigenstates are bound states. Eigenfunctions and eigenvalues of 
$D_a^\dagger D_a d_{nl} = m_{D,nl}^2 d_{nl}$ are given by
\be
d_{nl} &=& \sqrt{\dfrac{2 ^{|l| + 1} (n - |l|) !}{\pi ( n ! )^{3}}}\, \Omega ^{|l| + 1} r ^{|l|} L^{|l|}_{n} \left( 2\Omega ^{2} r^{2} \right) e^{-\Omega ^{2} r^{2} + i l \phi}, \\
L^{|l|}_{n} ( \xi ) &\equiv & \dfrac{d ^{\, |l|}}{d \xi ^{\, |l|}} e ^{\, \xi} \dfrac{d ^{\, n}}{d \xi ^{\, n}} \xi ^{\, n} e ^{- \xi}, \\
m_{D,nl} &=& 2 \sqrt{2n - |l|} \, \Omega,
\label{eq:bbbb}
\ee
with $n$ is semi-positive integer and $l$ is an integer of $|l|\le n$.
$L^{|l|}_{n}$ is the associated Laguerre polynomials.
The zero mode is unique with $(n,l) = (0,0)$ with the wave function
\be
d_{00} \propto e^{-\Omega^2 r^2}.
\ee

The mass spectrum of the divergence-free part 
$A_a^{\widehat{\rm df}}$ is determined by ${\cal H}$ from Eq.~(\ref{eq:extraA_Lag_geneD}).
We have shown in Sec.~\ref{sec:A_a_neutral_D} 
that the mass spectrum of ${\cal H}$ is identical to that of $\bar{\cal H}$.
In six dimensions, $\bar{\cal H}$ is especially simple as
\be
\bar{\cal H} = {\bm D}{\bm D}^\dagger = D_aD_a^\dagger = D^\dagger_a D_a + 8\Omega^2.
\ee
Therefore, the mass spectrum of $A_a^{\rm df}$ is given by
\be
\bar m_{D,nl}^2 = m_{D,nl}^2  + 8\Omega^2.
\label{eq:aaaa}
\ee

\subsubsection{Charged stablizer}

Next we consider the case that the stabilizer $H$ is charged.
The masses of $A_\mu$ are identical to eigenvalues of the operator $\mathsf{D}_a^\dagger\mathsf{D}_a
= D_a^\dagger D_a + q^2{\cal E}_0^2$.
For $\beta$ in Eq.~(\ref{eq:beta_D6_2}), ${\cal E}_0 = \sqrt{2}\,q\mu$ is constant. So, the
effect of changing the neutral $H$ by the charged $H$ is just shift of all the eigenvalues
$m_{D,nl}^2$ of $D_a^\dagger D_a$ by the constant $2q^2\mu^2$. This is the consequence of the local
Higgs mechanism.

According to the generic arguments in Sec.~\ref{sec:Charged_Stab_D}, the other physical 
degrees of freedom live in the divergence-free part $\vec{\mathsf{A}}^{\widehat{\rm df}}$.
It can be further decomposed to two components orthogonal to each other:
the component associated with $\bar{f}$  
and the other component associated with $\vec{A}^{\rm d}$ of (\ref{eq:Adf_sf}).
The former has the eigenvalues of $\bar {\cal H} + 2 q^2\mu^2$, and the latter has 
the eigenvalues of $\vec D \vec D^\dagger + 2q^2\mu^2{\bf 1}_2$. 
In the six dimensions, we have $\bar{\cal H} = D_a D_a^\dagger$ and its eigenvalues 
are given in Eq.~(\ref{eq:aaaa}).
On the other hand, the non-zero eigenvalues of 
$\vec D \vec D^\dagger$ and $\vec D^\dagger \vec D = D_a^\dagger D_a$ are always identical,
which is given in Eq.~(\ref{eq:bbbb}).
Therefore, the mass spectra in $\vec{\mathsf{A}}^{\widehat{\rm df}}$ are
$m_{D,nl}^2 + 8\Omega^2 + 2q^2\mu^2$ and $m_{D,nl}^2 + 2q^2\mu^2$ with $(n,l) \neq (0,0)$.

\section{Conclusion}
\label{sec:conclusion}

In this paper we investigated localization of the gauge fields via the field dependent
gauge kinetic term (\ref{eq:f2}) in details. We considered two cases that the stabilizers
are neutral and charged. For the neutral case, we improved previous analysis done in
Ref.~\cite{Ohta:2010fu,Arai:2012cx,Arai:2013mwa,Arai:2014hda,Arai:2016jij,Arai:2017lfv,Arai:2017ntb},
and especially analysis on the divergence-free parts becomes better, see Sec.~\ref{sec:2} for
$D=5$, and Sec.~\ref{subsec:neutral_D} for generic $D\ge 5$.
We also studied the models with charged stabilizers. The charged stabilizers are locally condensed
inside a topological soliton, so that they localize the gauge fields and at the same time
they give a finite mass to the localized gauge fields similarly to the conventional Higgs mechanism.
In order to determine physical mass spectra, we need to diagonalize complicated mixings between 
the four-dimensional gauge fields, Nambu-Goldstone fields, and the extra-dimensional gauge fields
which are further decomposed into the divergence and divergence-free parts.
We developed complete and self-contained formula with which one can clearly separate physical
and unphysical degrees of freedom for generic models in generic $D\ge5$ dimensions.

When we want massless gauge fields on a topological soliton, all we have to do is 
preparing neutral stabilizers which interact with the would-be localized gauge fields
via Eq.~(\ref{eq:f2}). On the other hand, if we want to have massive gauge bosons
on a topological soliton, it can be realized by just replacing the neutral stabilizers with
the charged ones. An attempt of identifying the charged stabilizer with the SM
Higgs boson in the $D=5$ model was studied in Ref.~\cite{Arai:2018uoy}.

Let us make a comment on localized gauge fields on domain walls in the Higgs vacua 
\cite{Tong:2002hi,Shifman:2002jm,Shifman:2003uh,Isozumi:2004jc,Isozumi:2004vg,Isozumi:2004va,Eto:2004vy,Tong:2005un,Eto:2006pg,Shifman:2007ce}.
When a domain wall interpolates two discrete Higgs vacua where the gauge symmetry is broken,
the gauge symmetry is approximately recovered inside the domain wall. Nevertheless, the localized
gauge field on the domain wall never becomes massless. The lowest mass of localized gauge field
is inevitably of order inverse of the domain wall that is the same order of all the KK modes. Therefore,
in principle, we cannot distinguish the lightest massive gauge bosons from the other KK modes.
In contrast, the solitons in this work live in the confining vacua where the gauge symmetry
is not broken. As we shown in this work, 
the lightest mass of localized gauge bosons is controlled only by the charge
of the stabilizer, and it is nothing to do with the soliton width. Therefore, we can
introduce two independent mass scales: the one is the lightest gauge boson mass and the
other is superheavy mass of the KK towers. For phenomenological purpose, see for example
Ref.~\cite{Arai:2018uoy}, our model has an
advantage compared to other models with topological solitons in the Higgs phase.

As a future direction, it might be interesting to study the intersections of domain walls
studied in Sec.~\ref{sec:example}. In this work, we only considered the intersection of
two domain walls at right angle. In general, we can consider multiple domain walls, for example,
three domain walls intersect at three different intersection points. We can also include 
fermions, and investigate whether such model provides a realist four dimensional model
like intersecting D-branes \cite{Higaki:2005ie}.


\section*{Acknowledgements}

M.\ E. thanks to Masato Arai, Filip Blaschke, and Norisuke Sakai for fruitful discussions
throughout long collaboration,
and contributions at the early stage of this work.
The work is supported in part by JSPS Grant-in-Aid for Scientific Research 
KAKENHI Grant No. JP16H03984 and No. JP19K03839.
The work is also supported in part by MEXT KAKENHI Grant-in-Aid for 
Scientific Research on Innovative Areas
Discrete Geometric Analysis for Materials Design No. JP17H06462 from the MEXT of Japan.

\bibliographystyle{jhep}

\end{document}